\documentclass[10pt,journal]{IEEEtran}
\IEEEoverridecommandlockouts

\usepackage{graphicx}
\usepackage{amsmath,amsthm,amsfonts,amssymb}
\usepackage{cite}
\usepackage{bm}
\usepackage{bbm}
\usepackage{url}
\usepackage{array}
\usepackage{color}
\usepackage{multirow}
\usepackage{booktabs}
\usepackage[table,xcdraw]{xcolor}
\usepackage{enumitem}
\usepackage{subcaption}

\usepackage[english]{babel}
\usepackage{algorithm}
\usepackage[noend]{algpseudocode}
\usepackage{setspace,soul}

\theoremstyle{plain}
\newtheorem{thm}{Theorem}
\newtheorem{lem}[thm]{Lemma}

\newtheorem{prop}[thm]{Proposition}
\newtheorem{cor}[thm]{Corollary}
\newtheorem{rem}{Remark}
\newtheorem{sty1}{Theorem}
\newtheorem{defi}[sty1]{Definition}

\newenvironment{NewProof}{{\noindent\it Proof.}}{\hfill $\blacksquare$\par}

\usepackage[english]{babel}
\usepackage{algorithm}
\usepackage[noend]{algpseudocode}

\allowdisplaybreaks[4]

\begin{document}
\title{DEEP-IoT: Downlink-Enhanced Efficient-Power Internet of Things}

\author{
Yulin~Shao,~\IEEEmembership{Member,~IEEE}
\thanks{Y. Shao is with the State Key Laboratory of Internet of Things for Smart City and the Department of Electrical and Computer Engineering, University of Macau, Macau S.A.R. (E-mail: ylshao@um.edu.mo).}
}

\maketitle

\begin{abstract}
At the heart of the Internet of Things (IoT) -- a domain witnessing explosive growth -- the imperative for energy efficiency and the extension of device lifespans has never been more pressing. This paper presents DEEP-IoT, an innovative communication paradigm poised to redefine how IoT devices communicate. Through a pioneering feedback channel coding strategy, DEEP-IoT challenges and transforms the traditional transmitter (IoT devices)-centric communication model to one where the receiver (the access point) play a pivotal role, thereby cutting down energy use and boosting device longevity.
We not only conceptualize DEEP-IoT but also actualize it by integrating deep learning-enhanced feedback channel codes within a narrow-band system. 
Simulation results show a significant enhancement in the operational lifespan of IoT cells -- surpassing traditional systems using Turbo and Polar codes by up to $52.71\%$. 
This leap signifies a paradigm shift in IoT communications, setting the stage for a future where IoT devices boast unprecedented efficiency and durability.
\end{abstract}


\section{Introduction}
The widespread application of the Internet of Things (IoT) has become a vital catalyst for societal, economic, and technological advancement \cite{IoT1,IoT_data,PAMA,LPWA}. From smart homes to industrial manufacturing, from smart healthcare to intelligent cities, IoT has permeated various sectors, providing key support for social transformation and technological progress. By connecting a vast array of devices, IoT not only facilitates the real-time collection of extensive data but also enables sophisticated data mining and analytical processes \cite{IoT_data,shao2020significant}. These capabilities provide a more accurate base for decision-making, thereby playing a crucial role in shaping innovative business models and strengthening the data backbone for scientific exploration and effective governance.

Amidst this backdrop, energy efficiency and device longevity stand out as pressing challenges. IoT devices are typically battery-powered with limited capacity, yet they are deployed over vast areas and are required to continuously update their sensing data to a remote access point (AP). The demand for extensive and deep coverage necessitates increased energy consumption for wireless transmission. For instance, in Narrowband IoT (NB-IoT) \cite{nbiot}, IoT devices have to repetitively transmit a data packet up to $128$ times to achieve a maximum path loss of $164$dB. Moreover, IoT devices often encounter challenges in timely energy replenishment. This becomes even more pronounced in critical use cases such as monitoring in deep wells or disaster surveillance in scenarios like typhoons or earthquakes, where battery replacement is not just challenging but sometimes impossible. The above factors underscore the importance of enhancing communication energy efficiency and extending the lifespan of IoT devices.

The predominant approach for reducing the communication power expenditure of IoT devices is to extend their sleep periods \cite{LPWA,nbiot,LoRa}. Take NB-IoT, for example, which implements two main power-saving strategies: power saving mode (PSM) and enhanced discontinuous reception (eDRX) \cite{nbiot}. In PSM, IoT devices transition into a partial sleep state after a specified duration of inactivity, significantly reducing the power expenditure of their radio chips. While in this mode, the devices remain registered on the network but cease listening for paging from the AP. On the other hand, eDRX allows devices to remain idle before entering PSM yet receptive to AP paging for some time, ensuring the reachability of downlink data. A similar concept exists in the LoRa IoT protocol with the Wake on Radio (WOR) mechanism \cite{LoRa}. WOR optimizes the communication energy expenditure while maintaining network connectivity by keeping IoT devices in a sleep state and periodically activating them to listen for any wake-up signals. Overall, prolonging the sleep cycles in NB-IoT and LoRa protocols has notably extended the lifetimes of IoT devices. Nevertheless, this method does not fundamentally modify the essence of communication, as it does not inherently increase the energy efficiency of the data transmission process itself: the energy expenditure for transmitting each unit of data to a remote AP remains the same, and the total amount of data an IoT device can transmit is still constrained by its power limitations.

This paper explores a pioneering energy-saving transmission strategy, dubbed Downlink-Enhanced Energy-Efficient IoT (DEEP-IoT), by innovating the channel coding techniques used in IoT wireless communications. The genesis of DEEP-IoT is rooted in analyzing the energy consumption patterns of IoT devices during transmission. Taking NB-IoT as an example, a typical NB-IoT chip's transmit power, receive power, and sleep power are approximately $135$mA, $4$mA, and $4\mu$A, respectively. Traditionally, the transmitter, possessing information unknown to the receiver, plays a dominant role in communications -- it transmits data continuously while the receiver passively responds with a $1$-bit acknowledgement (ACK) or negative ACK (NAK). In this setup, the major energy consumption occurs at the transmitting end.

However, recognizing that 1) the transmit power consumption of IoT devices far exceeds their receive power consumption, 2) the AP has a comparatively more abundant energy supply than IoT devices, we advocate for a fundamental paradigm shift: empowering the receiver with a more proactive role by enabling it to provide real-time feedback on its decoding status to the transmitter. With access to this feedback, the transmitter can dynamically adjust its coding strategy to match the receiver's current conditions. This iterative process of feedback and adaptation allows for a more efficient use of the transmitter's energy, embodying a ``listen more, transmit less'' approach -- by listening more feedback than traditional systems from the AP, DEEP-IoT can transmit the same amount of information bits as traditional systems but at a significantly lower power.
Adopting this strategy, DEEP-IoT essentially facilitates an energy exchange between IoT devices and the AP, leading to a marked decrease in energy consumed for transmissions and a notable extension of the devices' operational lifespan.

The significance of this paper are as follows:
\begin{itemize}
\item We challenge and move beyond the traditional paradigm of transmitter-dominated, unidirectional communication by introducing an innovative IoT communication framework: DEEP-IoT. This approach, distinct from and complementary to existing energy-saving strategies like PSM and eDRX, offers a pathway toward achieving exceptionally low power consumption for IoT communications.
\item We introduce the inaugural system design of DEEP-IoT, integrating it within a narrow-band single
carrier-frequency division multiple access (SC-FDMA) framework. This design incorporates advanced deep learning techniques to optimize feedback channel codes, accommodating extensive feedback scenarios not previously addressed. We establish a methodological approach to empirically model the relationship between the required signal-to-noise ratio (SNR) and feedback volume, utilizing a logistic function to predict and manage energy efficiency based on dynamic network conditions.
\item At the multiple-access layer, DEEP-IoT leverages a half-duplex frequency division duplex (HD-FDD) configuration, leading to a new dual-channel access framework. We develop a novel index table-based policy gradient method, which simplifies the complexity of optimizing feedback distribution across multiple devices. This strategic framework significantly enhances the longevity of an IoT cell -- by $52.71\%$ and $42.34\%$ over traditional systems utilizing Polar and Turbo codes, respectively.
\item Our comprehensive system design and empirical evaluations confirm that DEEP-IoT not only meets but significantly exceeds traditional performance benchmarks, providing a robust, scalable solution that adapts to the evolving demands of IoT environments.
\end{itemize}

The remainder of this paper is organized as follows.
Section~\ref{sec:II} reviews existing literature and developments in feedback channel coding.
Section~\ref{sec:III} outlines the framework of our DEEP-IoT system, introducing its innovative approach to IoT communications.
Sections~\ref{sec:IV} and \ref{sec:V} are dedicated to the in-depth exploration of the DEEP-IoT system design, covering the physical layer and multiple-access layer, respectively. These sections also include simulation results that illustrate the effectiveness of our design.
Sections~\ref{sec:Conclusion} concludes this paper.

{\it Notations}: We use boldface letters to denote column vectors (e.g., $\bm{x}$, $\bm{y}$). For a vector or matrix, $(\cdot)^\top$ denotes the transpose, $(\cdot)^*$ denotes the complex conjugate, and $(\cdot)^H$ denotes the conjugate transpose. $\mathbb{R}$ and $\mathbb{C}$ stand for the sets of real and complex numbers, respectively. The imaginary unit is represented by $j$. $\mathcal{N}$ and $\mathcal{CN}$ stand for the real and complex Gaussian distributions, respectively. The cardinality of a set $\mathcal{A}$ is denoted by $|\mathcal{A}|$. $\text{sgn}(\cdot)$ denotes the sign function; $\delta(\cdot)$ denotes the Dirac delta function.

\section{Related Work}\label{sec:II}
% {\it Related work} -- 
Feedback is an indispensable element in communications \cite{Shannon,lovefeedback,Polyanskiy,DeepCode}. In modern systems, where the transmitter typically takes the lead, the receiver's feedback is only used to confirm the successful receipt of messages. A notable example is Hybrid Automatic Repeat reQuest (H-ARQ), which uses 1-bit feedback to prompt the transmitter to send additional parity symbols. This allows the receiver to combine multiple transmissions to improve decoding performance. From this perspective, DEEP-IoT can be viewed as an advanced iteration of H-ARQ that enhances performance by providing more comprehensive feedback about the AP's decoding status.

In 1956, C. E. Shannon set a cornerstone by introducing the classical feedback channel model \cite{Shannon}, in which transmitter and receiver are connected through a forward channel and a feedback channel. The forward channel is characterized as a discrete-time, memoryless additive white Gaussian noise (AWGN) channel, while the feedback channel is noiseless with a unit delay -- every piece of information received by the receiver will be instantly fed back to the transmitter, providing the later with immediate insights on the noise realization encountered in the forward transmission. Although Shannon proved that feedback does not increase the classic channel capacity of the forward channel, it does enhance the reliability of forward transmissions in the case of finite block lengths. Building on this, in 1967, Schalkwijk and Kailath introduced the seminal SK coding scheme within Shannon’s feedback channel model \cite{SK1,SK2}. The SK scheme involves pulse-amplitude modulation (PAM) and iterative residual error transmission to refine the message estimate at the receiver, achieving a double exponential decrease of decoding error probability in the block length, provided that the transmission rate is below channel capacity.

Despite the impressive performance, the SK coding scheme is vulnerable to even minimal noise in the feedback channel. This vulnerability has spurred interest in developing feedback coding strategies suited for imperfect feedback conditions. In \cite{Kim2007ISIT}, Y. H. Kim et al. highlighted the futility of linear codes in enhancing communication rates under noisy feedback, suggesting instead a nonlinear approach. This alternative method employs a three-stage process of detection and retransmission, which proved to be nearly optimal. On the other hand, A. Ben-Yishai and O. Shayevitz reevaluated the SK scheme in \cite{MSK} and treated feedback coding as a joint source-channel coding (JSCC) problem with side information when faced with noisy feedback. Given this insight, they improved the SK scheme with modulo operations and developed the modulo SK (MSK) scheme, which demonstrates superior performance, particularly when the feedback channel's signal-to-noise ratio (SNR) significantly outperforms that of the forward channel.

Another category of feedback codes is derived from deep learning (DL), by learning from data. A pioneering attempt in this direction was made by DeepCode \cite{DeepCode}. Inspired by \cite{Kim2007ISIT}, the authors of DeepCode believed that nonlinear coding structures play a crucial role in the design of feedback codes, while deep neural networks (DNNs) are renowned for their robust nonlinear fitting capabilities. In the DL-based feedback code design, the communication system is treated as an autoencoder, with both the feedback encoder and decoder parameterized as DNNs. By simulating the data generation and transmission processes, the parameters within the autoencoder are trained to minimize the end-to-end bit error rate (BER). Ultimately, when the learning process culminates with minimized BER, effective feedback encoder and decoder structures are derived. In DeepCode, the authors utilized recurrent neural networks (RNNs) as its encoding and decoding backbones and achieved commendable error performance under Shannon's classic feedback channel model, demonstrating the feasibility of designing feedback codes via DL.

Following the footsteps of DeepCode, AttentionCode \cite{AttentionCode} introduced a feedback encoding scheme based on the self-attention mechanism for the classic feedback channel model. Its backbone, dubbed AttentionNet, is derived from the increasingly dominant transformer architecture \cite{transformer,Bert}. Despite the added complexity in training, AttentionCode achieved significantly better coding performance than DeepCode. Leveraging the encoding capabilities of AttentionNet and the block-wise attention mechanism inspired by Vision transformers, the authors of \cite{GBAFC} proposed a new class of feedback codes: generalized block attention feedback (GBAF) code. The significance of GBAFC code lies in transcending the classic feedback channel model, shifting feedback code design from symbol-based to packet-based feedback. This development laid the groundwork for the feedback framework between IoT devices and APs in DEEP-IoT. Subsequent work \cite{active} further expanded the concept of passive feedback at the receiver end to active feedback, which proved to be highly effective in noisy feedback channels.

Overall, DL's ability to handle complex, non-linear tasks makes it ideally suited for modern communication challenges. These challenges often involve intricate and dynamic environments, which traditional designs struggle to address effectively. By leveraging DL, we can develop more robust feedback codes that can operate efficiently in diverse and evolving channel conditions, which is crucial for the advancement of technologies like IoT.

\section{DEEP-IoT System Overview}\label{sec:III}
\begin{figure}[t]
  \centering
  \includegraphics[width=0.9\columnwidth]{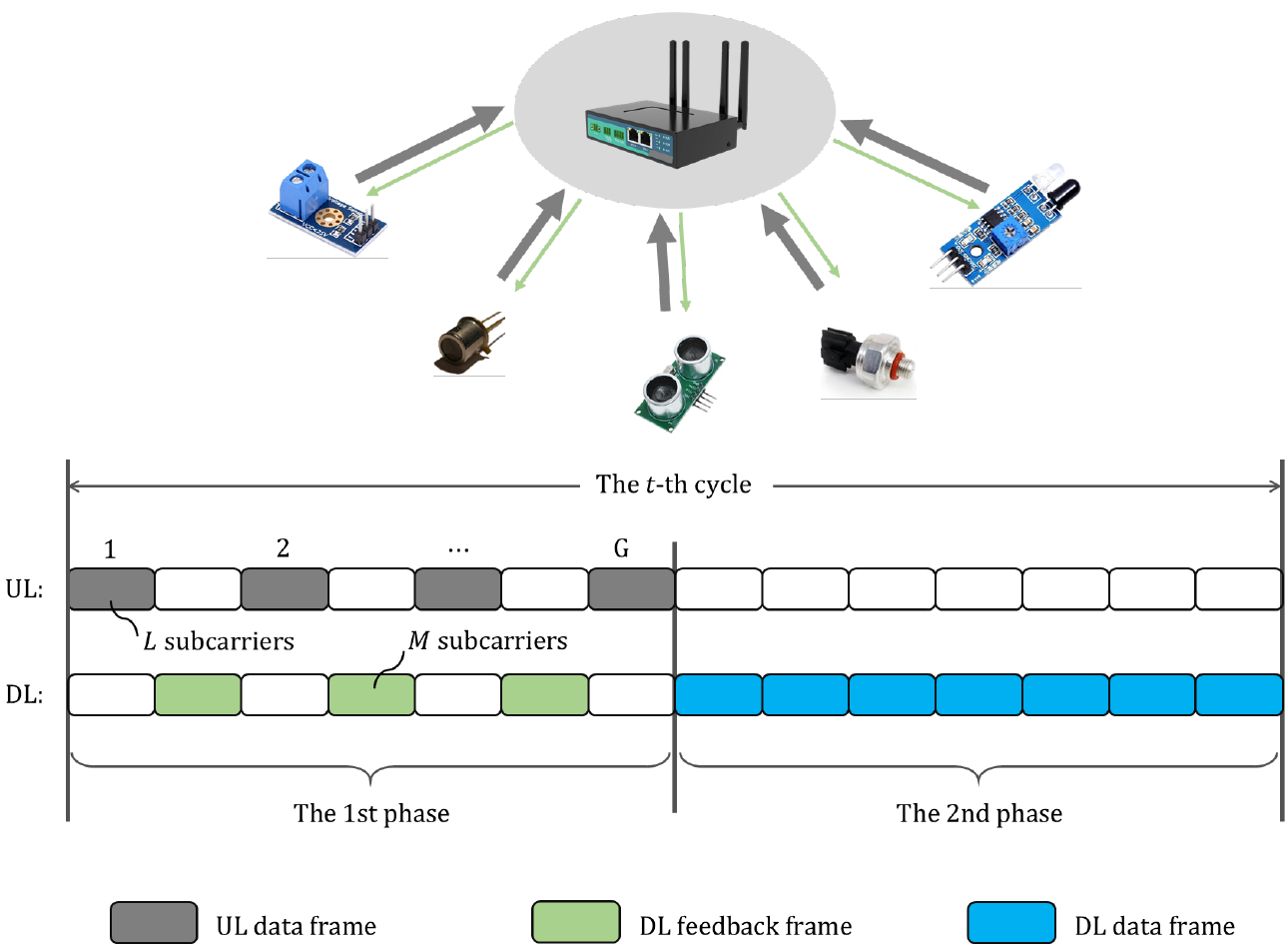}\\
  \caption{The system model and duplexing mode of DEEP-IoT.}
\label{fig:sys}
\end{figure}

We consider an IoT cell featuring an AP and $L$ distributed IoT devices deployed for environmental sensing, as depicted in Fig.~\ref{fig:sys}. DEEP-IoT operates in the HD-FDD mode. Time is segmented into cycles, denoted by $t=0,1,2,\cdots$. In each cycle, IoT devices are required to transmit their sensing data, quantified in $K$ bits, to the AP. Each cycle is structured into two phases: the 1st phase for uplink data transmission from the devices, and the 2nd phase for downlink data transmission from the AP. A key differentiator of DEEP-IoT from traditional communication systems is its approach to uplink data transmission. Hence, this paper primarily concentrates on the 1st phase of each cycle.

Without loss of generality, we focus on the 1st phase of the $t$-th cycle, and details the process by which devices transmit their data to the AP. As illustrated in Fig.~\ref{fig:sys}, DEEP-IoT divides the 1st phase into $2G-1$ sequential data frames, with $G$ frames allocated for uplink data transmission, termed UL data frames, and the remaining $G-1$ frames for downlink feedback, termed DL feedback frames. As can be seen, this interspersing of DL feedback frames amongst UL data frames is a distinctive feature that sets DEEP-IoT apart from legacy communication systems. The purpose of the AP sending these feedback frames to the IoT devices is to enable them to ascertain the current decoding status of the AP. This knowledge allows the devices to modify their encoding schemes for subsequent transmissions. Such a mechanism significantly enhances the uplink channel coding efficiency. As a result, the IoT devices can achieve the same coverage area and communication performance as traditional systems but with reduced power consumption.

We assume that each UL data frame and DL feedback frame comprises $Q$ orthogonal frequency-division multiplexing (OFDM) symbols. Specifically,
\begin{itemize} 
    \item The uplink data transmission is operated with SC-FDMA. Unless stated otherwise, we assume in this paper that each device occupies a single subcarrier, akin to the NB-IoT, to ensure sufficient coverage and signal penetration. The total count of subcarriers used for uplink data transmission is thus $L$. Adapting DEEP-IoT for cases where each device occupies multiple subcarriers is straightforward.
    \item The downlink feedback transmission is operated with OFDMA. We assume availability of $M$ subcarriers for downlink feedback, and each IoT device is assigned $a^{(t)}_{\ell}$ subcarriers, where $a^{(t)}_{\ell}\in\{0,1,2,\cdots,M\}$, $\ell=1,2,\cdots,L$.
\end{itemize}

Overall, for the transmission of $K$ bits in the 1st phase, each user sequentially transmits $G$ UL frames using the assigned subcarrier. The total count of real channel symbols each device transmits to the AP is $2QG$. In between two consecutive uplink frames, the AP feeds back a downlink frame. The total count of real symbols fed back to the $\ell$-th device is $2Q(G-1)a^{(t)}_{\ell}$.

\section{DEEP-IoT Physical Layer}\label{sec:IV}

\begin{figure*}[t]
  \centering
  \includegraphics[width=1.4\columnwidth]{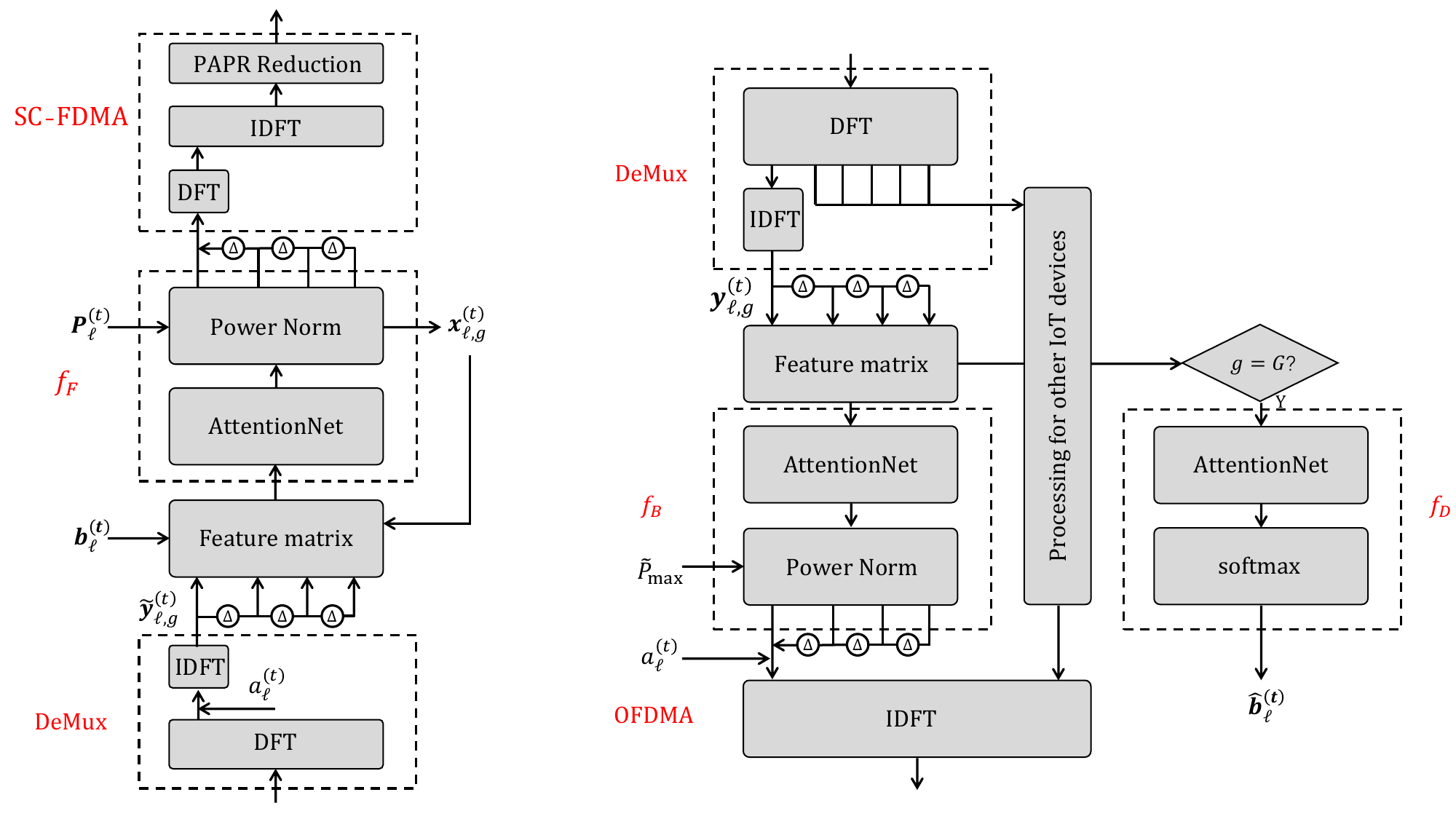}\\
  \caption{Feedback channel coding/decoding and modulation/demodulation at the IoT devices and the AP in DEEP-IoT.}
\label{fig:sys2}
\end{figure*}

In this section, we focus on the signal processing at the IoT devices and AP, and explain how DEEP-IoT manages channel coding/decoding and signal modulation/demodulation. This is crucial for grasping the nuances of DEEP-IoT's design and its potential advantages over conventional IoT systems.

\subsection{Signal flow}\label{sec:IVA}
Consider the 1st phase of the $t$-th cycle, in which the IoT devices sequentially transmits $G$ frames to the AP. For the $\ell$-th device, we denote by $\bm{x}^{(t)}_{\ell,g}$, $g=1,2,\cdots,G$, the vector of real channel symbols transmitted in the $g$-th frame. Since each device occupied one subcarrier, we have $\bm{x}^{(t)}_{\ell,g}\in\mathbb{R}^{2Q\times 1}$. After OFDM demodulation, the signal received at the AP can be written as
\begin{equation}\label{eq:III1}
\bm{y}^{(t)}_{\ell,g}=\sqrt{\alpha_{\ell}P^{(t)}_{\ell}} \left|h^{(t)}_{\ell}\right|\bm{x}^{(t)}_{\ell,g} + \bm{w}^{(t)}_{\ell,g},
\end{equation}
where $\alpha_{\ell}$ is the path loss coefficient, $P^{(t)}_{\ell}\leq P_{\max}$ is the average UL transmit power, $h^{(t)}_{\ell}\sim\mathcal{CN}(0,\sigma^2_{\ell})$ is the channel gain associated with the subcarrier assigned to the $\ell$-th device, the power of $\bm{x}^{(t)}_{\ell,g}$ has been normalized to $1$, and $\bm{w}^{(t)}_{\ell,g}$ is the additive white Gaussian noise (AWGN) vector with a power spectrum density of $N_0$. We consider slow fading channel, hence the channel gain $h^{(t)}_{\ell}$ is constant during the 1st phase of the $t$-th cycle. In particular, we assume that the channel state information is available at both devices and AP, and the phase of $h^{(t)}_{\ell}$ has been compensated \cite{shao2021federated} at the IoT device by a factor of $(h^{(t)}_{\ell})^*/\left|h^{(t)}_{\ell}\right|^2$. Therefore, we have $\left|h^{(t)}_{\ell}\right|$ in \eqref{eq:III1} and the UL received SNR is given by
\begin{equation}\label{eq:III2}
\eta^{(t)}_{\ell} \triangleq \frac{\alpha_{\ell}\left|h^{(t)}_{\ell}\right|^2}{N_0}P^{(t)}_{\ell}.
\end{equation}
In this paper, we will also write SNR in decibels, in which case a `dB' will be added in the subscript, for example, $\eta^{(t)}_{\ell,\text{dB}}\triangleq 10\lg \eta^{(t)}_{\ell}$.

Denote by $\widetilde{\bm{x}}^{(t)}_{\ell,g}$ the vector of real channel symbols fed back by the AP in the $g$-th DL feedback frame to the $\ell$-th IoT device. $\widetilde{\bm{x}}^{(t)}_{\ell,g}\in\mathbb{R}^{2Qa^{(t)}_{\ell}\times 1}$ is generated based on all the UL channel symbols previously received from the $\ell$-th device, giving
\begin{equation}\label{eq:III3}
\widetilde{\bm{x}}^{(t)}_{\ell,g}=f_B\left(\bm{y}^{(t)}_{\ell,1},\bm{y}^{(t)}_{\ell,2},\cdots,\bm{y}^{(t)}_{\ell,g} \right),
\end{equation}
where $f_B$ denotes the feedback encoder at the AP.

After passing through the feedback channel, the received signal at the $\ell$-th device can be written as
\begin{equation}\label{eq:III4}
\widetilde{\bm{y}}^{(t)}_{\ell,g}=\sqrt{\alpha_{\ell}\widetilde{P}_{\max}} \left|\widetilde{h}^{(t)}_{\ell}\right|\widetilde{\bm{x}}^{(t)}_{\ell,g} + \widetilde{\bm{w}}^{(t)}_{\ell,g},
\end{equation}
where the path loss coefficient $\alpha_{\ell}$ is the same as UL, $\widetilde{P}_{\max}$ is the maximum transmit power of the AP, $\widetilde{h}^{(t)}_{\ell}\sim\mathcal{CN}(0,\widetilde{\sigma}^2_{\ell})$ is the channel coefficient, and $\widetilde{\bm{w}}^{(t)}_{\ell,g}$ is the AWGN vector with a power spectrum density of $N_0$. The downlink feedback SNR can be written as
\begin{equation}\label{eq:III5}
\widetilde{\eta}^{(t)}_{\ell} \triangleq \frac{\alpha_{\ell}\left|\widetilde{h}^{(t)}_{\ell}\right|^2}{N_0}\widetilde{P}_{\max},
\end{equation}
or $\widetilde{\eta}^{(t)}_{\ell,\text{dB}}= 10\lg \widetilde{\eta}^{(t)}_{\ell}$ in decibels.

Given the feedback from the AP, the $\ell$-th user generates the $(g+1)$-th uplink packet $\bm{x}^{(t)}_{\ell,g+1}$ based on all information bits, the previously transmitted coded symbols, and all symbols fed back from the AP, yielding
\begin{equation}\label{eq:III6}
\left\{
\begin{array}{l}
   \bm{x}^{(t)}_{\ell,1}=f_F\left(\bm{b}^{(t)}_{\ell}\right), \\
   \bm{x}^{(t)}_{\ell,g+1}=f_F\left(\bm{b}^{(t)}_{\ell};\bm{x}^{(t)}_{\ell,1},\cdots,\bm{x}^{(t)}_{\ell,g};\widetilde{\bm{y}}^{(t)}_{\ell,1},\cdots,\widetilde{\bm{y}}^{(t)}_{\ell,g}\right),
\end{array}
\right.
\end{equation}
where $\bm{b}^{(t)}_{\ell}\in\{0,1\}^{K\times 1}$ is the sensing data to be reported to the AP by the $\ell$-th device, and $f_F$ denotes the feedback encoder at the devices.\footnote{We point out that both feedback encoders $f_B$ and $f_F$ are no long channel encoders, but joint source-channel encoders.}

At the end of the 1st phase, the AP receives $G$ frames from each IoT device and decodes each device's data using a feedback decoder $f_D$:
\begin{equation}\label{eq:III7}
\widehat{\bm{b}}^{(t)}_{\ell}=f_D\left(\bm{y}^{(t)}_{\ell,1},\bm{y}^{(t)}_{\ell,2},\cdots,\bm{y}^{(t)}_{\ell,G} \right).
\end{equation}

Overall, in the 1st phase, each IoT device transmits $N\triangleq 2QG$ real channel symbols to the AP. The channel coding rate of each DEEP-IoT device, defined as the number of transmitted bits per coded channel symbol, is given by
\begin{equation}\label{eq:III8}
R \triangleq \frac{K}{N} = \frac{K}{2QG}.
\end{equation}

\begin{rem}
In this paper, we consider fixed-rate feedback codes for DEEP-IoT, meaning that the code rate remains constant throughout the encoding process \cite{DeepVLF}. The primary dynamic adjustment lies in the generation of parity symbols by the encoding network. Although the number of transmitted symbols remains fixed, improved coding efficiency allows for significant reductions in transmission power. This is achieved through channel estimation conducted at the beginning of each communication cycle, where the observed channel gains inform the tuning of transmission power. By dynamically adapting the transmission power in response to channel conditions, DEEP-IoT optimizes energy usage while maintaining reliable communication.
\end{rem}

\subsection{Feedback channel coding and modulation}\label{sec:IVB}
Based on the signal flow presented in Section \ref{sec:IVA}, this section further delves into the baseband operations within both the IoT devices and the AP. Our focus will be on elucidating the generation of $\widetilde{\bm{x}}^{(t)}_{\ell,g}$ and $\bm{x}^{(t)}_{\ell,g}$ in \eqref{eq:III3} and \eqref{eq:III6}. For a comprehensive understanding, Fig. \ref{fig:sys2} offers a detailed illustration of the entire process.

Notably, the feedback coding framework utilized in DEEP-IoT is built upon our prior research \cite{AttentionCode,GBAFC,active}.
Contrary to previous studies that predominantly focus on limited feedback scenarios, our work pioneers the extensive feedback model within the IoT communications. This extension is guided by the ``listen more, transmit less'' philosophy, which is particularly advantageous in IoT setups where the energy availability at the AP is substantial compared to the devices. 
On the other hand, unlike prior studies that assume noiseless or nearly noiseless feedback conditions, this paper addresses practical, very noisy feedback environments that are typical in real-world applications. 
In the following, our discussion will be centered on meticulously designing the physical layer to seamlessly integrate the feedback coding mechanism into the overall framework of DEEP-IoT. The detailed network architectures and training methods are presented in Appendix \ref{sec:AppD}.

\subsubsection{UL feedback coding}
Let us start from the UL feedback encoder $f_F$ at the IoT devices, the primary function of which is to generate the next UL data frame, as in \eqref{eq:III6} Without loss of generality, we consider the generation of the $(g+1)$-th data frame $\bm{x}^{(t)}_{\ell,g+1}$. At that point, the available information at the $\ell$-th device comprises the original information bits $\bm{b}^{(t)}_{\ell}$, the previously transmitted symbols $\{\bm{x}^{(t)}_{\ell,1},\cdots,\bm{x}^{(t)}_{\ell,g}\}$, and all symbols fed back from the AP $\{\widetilde{\bm{y}}^{(t)}_{\ell,1},\cdots,\widetilde{\bm{y}}^{(t)}_{\ell,g}\}$.

\begin{figure}[t]
  \centering
  \includegraphics[width=0.7\columnwidth]{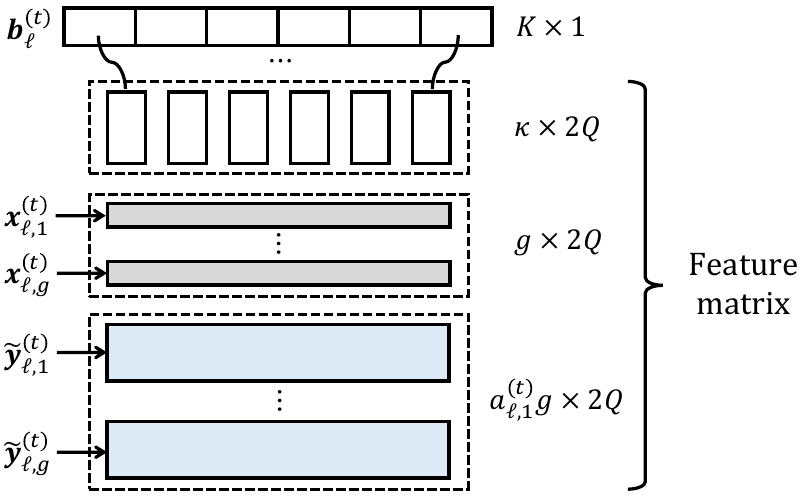}\\
  \caption{The composition of feature matrix $\bm{M}_F$ for UL feedback coding.}
\label{fig:FM}
\end{figure}

To perform UL encoding, we re-organize the above information into a feature matrix $\bm{M}_F$ that serves as the input for the UL feedback encoder $f_F$. As depicted in Fig.~\ref{fig:FM}, $\bm{M}_F$ is composed of three submatrices:
\begin{itemize}
    \item The first part consists of the original bits $\bm{b}^{(t)}_{\ell}$. We divide $K$ bits into $2Q$ groups, each containing $\kappa\triangleq K/2Q$ bits. Subsequently, $\bm{b}^{(t)}_{\ell}$ is reshaped into a $\kappa\times 2Q$ matrix, forming the first submatrix.
    \item The second part consists of previously transmitted channel symbols $\{\bm{x}^{(t)}_{\ell,1},\cdots,\bm{x}^{(t)}_{\ell,g}\}$. These are stacked together to form a $g\times 2Q$ submatrix.
    \item The third part consists of the received feedback $\{\widetilde{\bm{y}}^{(t)}_{\ell,1},\cdots,\widetilde{\bm{y}}^{(t)}_{\ell,g}\}$. Given that each $\widetilde{\bm{y}}^{(t)}_{\ell,g}$ is of dimension $2Qa^{(t)}_{\ell}\times 1$, we reshape these feedback symbols into an $a^{(t)}_{\ell}g\times 2Q$ submatrix.
\end{itemize}

The feature matrix is then fed into the UL feedback encoder $f_F$, which comprises three main components: a noise suppression module, a self-attention mechanism module, and a power normalization module. The detailed architecture of $f_F$ is given in our technical report \cite{DEEPIoT}. The output of $f_F$, $\bm{x}^{(t)}_{\ell,g}$, satisfies average power constraints:
    \begin{equation}\label{eq:III9}
        \frac{1}{N}\mathbb{E}\left[\sum_{g=1}^G \left(\bm{x}^{(t)}_{\ell,g} \right)^\top \bm{x}^{(t)}_{\ell,g} \right]\leq 1.
    \end{equation}
Then, based on the configured power $P^{(t)}_{\ell}$ of the $\ell$-th IoT device, we scale $\bm{x}^{(t)}_{\ell,g}$ by $\sqrt{P^{(t)}_{\ell}}$ for transmission, as in \eqref{eq:III1}.

\subsubsection{SC-FDMA}
In the $g$-th UL data frame, the $2Q$ coded symbols $\bm{x}^{(t)}_{\ell,g}$ are modulated onto $Q$ OFDM symbols using the designated subcarrier. As depicted in Fig.~\ref{fig:sys2}, the elements of $\bm{x}^{(t)}_{\ell,g}$ are sequentially fed into the SC-FDMA module, undergoing discrete fourier transform (DFT) and inverse DFT (IDFT). 
It is crucial to recognize that, when an IoT device occupies only a single subcarrier, the peak-to-average power ratio (PAPR) typically does not pose an issue. However, when devices occupy multiple subcarriers, a PAPR reduction module becomes necessary.

DEEP-IoT distinguishes itself by utilizing discrete-time continuous amplitude signals instead of traditional quadrature amplitude modulation (QAM) symbols. This affords the system greater flexibility in designing more efficient feedback encoders. At the AP's end, the feedback decoder $f_D$ will directly decode the likelihood ratio of the transmitted bits from the received continuous amplitude signals.
In the context of such discrete-time analog transmission (DTAT) systems, our previous research \cite{DTAT} has demonstrated that clipping with retraining is a highly effective PAPR reduction technique. This approach ensures that the system achieves a PAPR performance on par with conventional QAM-based SC-FDMA while minimizing any substantial degradation in overall performance. 
A thorough evaluation of DEEP-IoT's PAPR performance falls outside the scope of this paper and remains an area for future exploration.

\subsubsection{DL feedback coding and OFDMA}
The feature matrix of the AP can be written as
\begin{equation}\label{eq:III10}
\bm{M}_B\triangleq\left[\bm{y}^{(t)}_{\ell,1},\bm{y}^{(t)}_{\ell,2},\cdots,\bm{y}^{(t)}_{\ell,g} \right].
\end{equation}
Compared with the feature matrix $\bm{M}_F$ at the IoT devices, $\bm{M}_B$ contains only the UL received data symbols. Although it is possible to include DL coded symbols in $\bm{M}_B$, our evaluation does not show any substantial benefits from this addition.

The feature matrix $\bm{M}_B$ is then fed into the DL feedback encoder $f_B$, which is architecturally similar to $f_F$ and includes a noise suppression module, a self-attention mechanism module, and a power normalization module. Two distinctions between $f_F$ and $f_B$ are:
\begin{itemize}
    \item The output dimension of $f_B$ is $a^{(t)}_{\ell}\times 2Q$, with the post-normalization symbols $\widetilde {\bm{x}}^{(t)}_{\ell,g}$ adhering to the power constraint:
     \begin{equation}\label{eq:III11}
        \frac{1}{a^{(t)}_{\ell}N}\mathbb{E}\left[\sum_{g=1}^G \left(\widetilde{\bm{x}}^{(t)}_{\ell,g} \right)^\top \widetilde{\bm{x}}^{(t)}_{\ell,g} \right]\leq 1.
    \end{equation}   
    \item In contrast to IoT devices that are power-limited, the AP has stable energy supplies. Therefore, the transmission power of the AP is fixed to $\widetilde{P}_{\max}$, as in \eqref{eq:III4}. 
\end{itemize}

The DL coded symbols $\widetilde{\bm{x}}^{(t)}_{\ell,g}$ will be transmitted to IoT devices in the $g$-th DL feedback frame. DEEP-IoT adopts OFDMA in the DL, whereby the code symbols for the $\ell$-th device are sequentially modulated onto $Q$ OFDM symbols using the assigned $a^{(t)}_{\ell}$ subcarriers. Upon receiving the DL feedback frame, each device extracts its respective data from the designated subcarriers and integrates them into its feature matrix $\bm{M}_F$, as depicted in Fig.~\ref{fig:FM}.

\subsubsection{Decoding}
The current cycle concludes upon completion of the $G$-th UL data frame. Each IoT device, by this point, has transmitted $N=2QG$ coded symbols to the AP, all of which are stored in the feature matrix for decoding by the AP. As shown in Fig.~\ref{fig:sys2}, the DNN design of the decoder $f_D$ is akin to that of $f_B$, except for the replacement of the power normalization module with a softmax function.
Recall that the bit sequence $\bm{b}^{(t)}_{\ell}$ has been segmented into $2Q$ groups on the IoT device side.
Mirroring the approach of QAM, the AP engages in joint decoding of the bits within each group rather than individual bit decoding \cite{GBAFC}.
This approach effectively transforms the binary classification problem to a $2^{\kappa}$-ary classification problem.
Specifically, denote by $\bm{Y}\triangleq\{Y_{ij}\}\in\mathbb{R}^{2^\kappa\times 2Q}$ the output matrix of AttentionNet within the decoder $f_D$, the loss function can be written as
\begin{equation}\label{eq:III12}
\mathcal{L}=\mathbb{E} \left[
\sum_{j=0}^{2Q-1}\left(
-\sum_{i=0}^{2^\kappa-1}
\delta(j=c_i)\log\frac{e^{Y_{ij}}}{\sum_{i=0}^{2^\kappa-1}e^{Y_{ij}}}
\right)
\right],
\end{equation}
where the expectation is taken over a batch of bit sequences, and $c_i$ is the ground truth of a group of bits, i.e., $c_i=\bm{b}^\top[2^{\kappa}-1,2^{\kappa}-2,\cdots,2^0]^\top$ for $\bm{b}\in\{0,1\}^{\kappa\times 1}$.

\begin{table}[t]
\renewcommand*{\arraystretch}{0.9}
    \caption{Parameter settings for DEEP-IoT.}
    \setlength{\tabcolsep}{5mm} % length of the table.
    \label{tab:params}
    \centering
\begin{tabular}{@{}ccc@{}}
\toprule
\textbf{Parameter} & \textbf{Description} & \textbf{Value} \\
\midrule
$K$       & \#bits per IoT device            &   $48$    \\
$N$       & \#channel uses per IoT device            &   $144$    \\
$R$          &  Channel coding rate           &  $1/3$     \\ 
$\kappa$          &  \#bits per group           &  $3$     \\ 
$G$          &  \#UL frames per cycle           &  $9$     \\ 
$Q$          &  \#OFDM symbols per frame           &  $8$     \\ 
$M$          &  \#DL subcarriers           &  $4$     \\
\midrule
$\alpha$  &  Path loss coefficient  &   $10^{-5}$    \\ 
$\sigma^2$  &  UL channel fading  &   $5$    \\ 
$\widetilde{\sigma}^2$  & DL channel fading   &  $5$     \\ 
$P_{\max}$  & UL maximum transmit power   &   $0.5A$    \\ 
$\widetilde{P}_{\max}$  & DL transmit power   & $4A$     \\ 
$N_0$  &  AWGN power density  &   $-90$dBm   \\ 
\bottomrule
\end{tabular}
\end{table}
% $10^{-3}A$

\subsection{Performance}
This section evaluates the DEEP-IoT physical layer's effectiveness in enhancing communication reliability and minimizing energy expenditure of IoT devices. To this end, we shall focus on the communications between the AP and a single IoT device, and compare the communication energy expenditure with and without the integration of feedback.

\subsubsection{PER}
Unlike traditional communication systems, DEEP-IoT operates on an interactive communication paradigm, where its system performance is influenced by both UL and DL channels. Therefore, our initial step involves evaluating the packet error rate (PER) performance of the DEEP-IoT physical layer under various conditions of UL channel SNR, DL channel SNR, and the number of available downlink subcarriers for feedback. 

\begin{figure*}[t]
    \centering
    \begin{subfigure}{.32\textwidth}
        \centering
        \includegraphics[width=\textwidth]{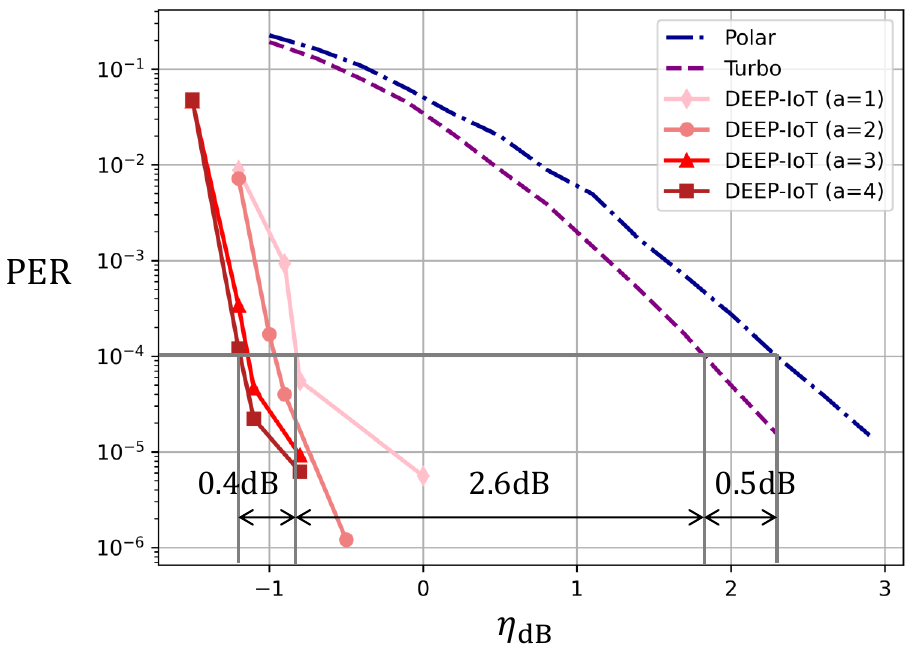}
        \caption{$\widetilde{\eta}_{\text{dB}}=20$dB}
    \end{subfigure}
    \begin{subfigure}{.32\textwidth}
        \centering
        \includegraphics[width=\textwidth]{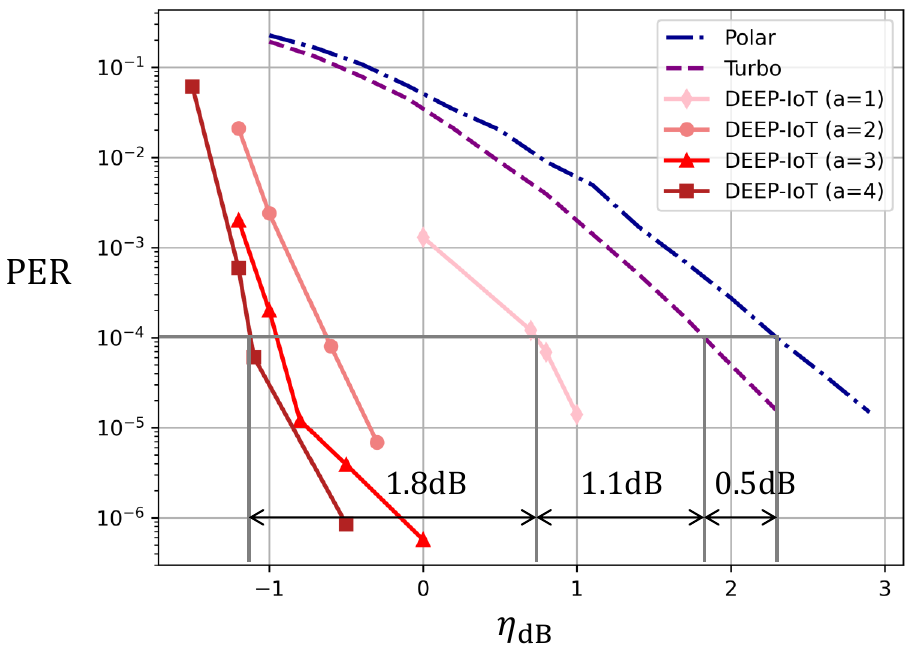}
        \caption{$\widetilde{\eta}_{\text{dB}}=10$dB}
    \end{subfigure}
    \begin{subfigure}{.32\textwidth}
        \centering
        \includegraphics[width=\textwidth]{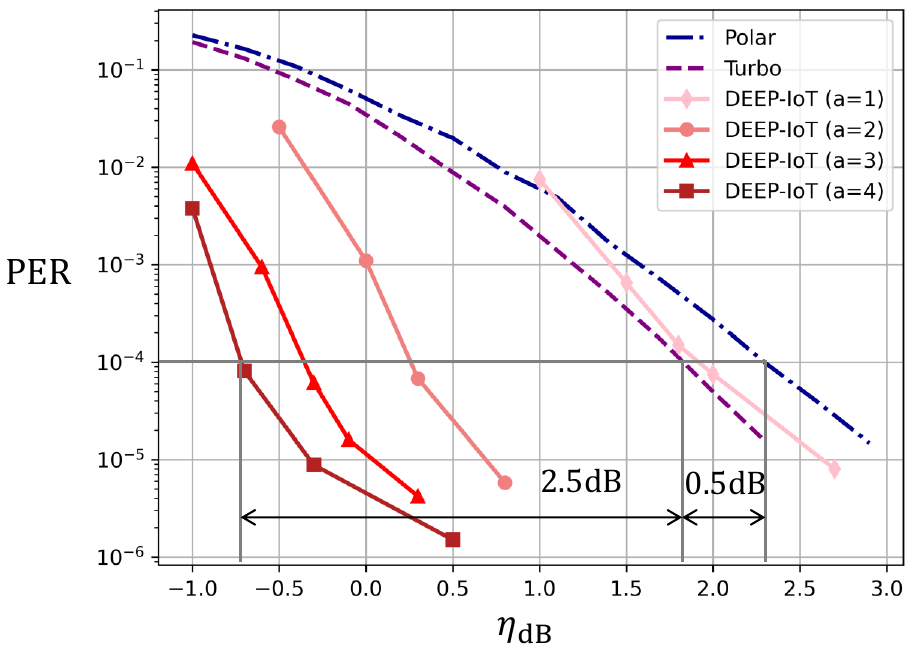}
        \caption{$\widetilde{\eta}_{\text{dB}}=5$dB}
    \end{subfigure}
    \caption{The PER performance of DEEP-IoT physical layer benchmarked against Polar code and Turbo code under various feedback SNR and available number of subcarriers for feedback.}
    \label{fig:S0}
\end{figure*}

We consider the simulation setup and parameter settings of DEEP-IoT in Table \ref{tab:params}. As we are examining a single IoT device's interaction with the AP in one cycle, the superscript $(t)$ and the subscript $\ell$ of all variables has been omitted. 
As shown in Table \ref{tab:params}, the IoT device is configured to upload $K=48$ bits of data per cycle, with $N=2QG=144$ available real channel uses. This results in a coding rate of $R=1/3$ for the IoT device.
The numerical results are presented in Fig.~\ref{fig:S0}, where the feedback channel SNR is set at $20$dB, $10$dB, and $5$dB, respectively. 
As baselines, we consider $(K,N)$ Turbo and Polar forward channel codes.

\begin{figure}[t]
  \centering
  \includegraphics[width=0.8\columnwidth]{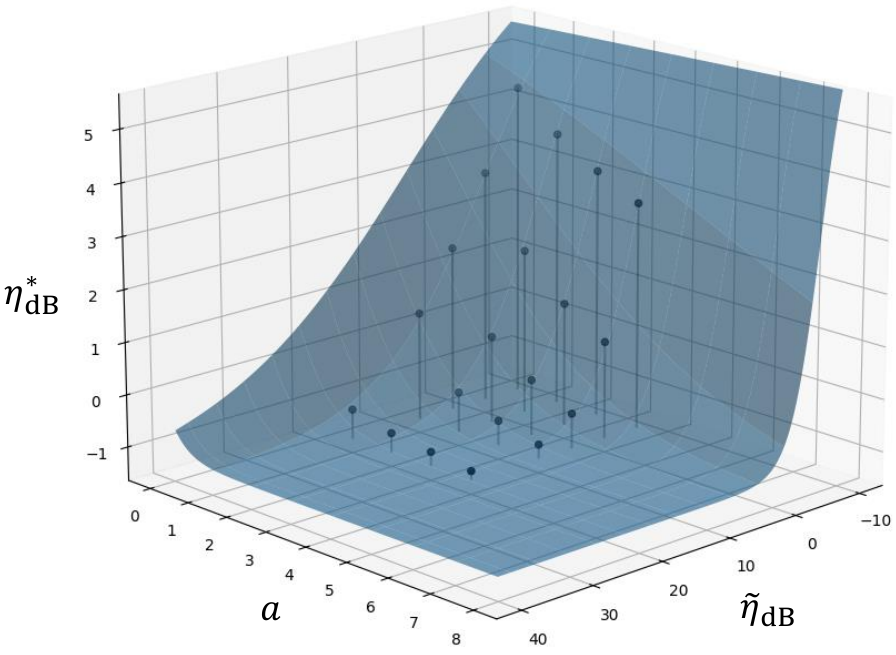}\\
  \caption{Illustration of the logistic function approximation for the required UL SNR in DEEP-IoT to meet a target PER, highlighting the dependency on DL feedback channel SNR and the allocation of feedback subcarriers to the IoT device, as defined in \eqref{eq:fit}.}
\label{fig:FS3D}
\end{figure}

As shown in Fig.~\ref{fig:S0}, DEEP-IoT demonstrates superior performance compared to traditional methods, significantly reducing the required uplink SNR to achieve the same PER. In particular, the PER performance of DEEP-IoT is closely linked to the quality of the DL channel. With a high DL channel SNR, AP can achieve remarkable PER gains even with just one feedback subcarrier i.e., the $a=1$ case in Fig.~\ref{fig:S0}. Increasing the number of feedback subcarriers does not substantially boost performance. For instance, with a DL SNR of $\widetilde{\eta}=20$dB and aiming for a PER of $10^{-4}$, DEEP-IoT with a single feedback subcarrier ($a=1$) surpasses the Polar code by $3.1$dB and the Turbo code by $2.6$dB. Increasing the number of feedback subcarriers to four only yields an additional $0.4$dB gain. Conversely, at a lower DL channel SNR, AP requires more feedback subcarriers to achieve significant coding gains. When the DL SNR is $\widetilde{\eta}=5$dB, for example, the UL SNR required by DEEP-IoT with one feedback subcarrier ($a=1$) to achieve a PER of $10^{-4}$ is on par with that required by Polar and Turbo codes. By increasing the number of DL feedback subcarriers to four, DEEP-IoT markedly outdoes the Polar and Turbo codes by $3$dB and $2.5$dB, respectively.

\begin{figure}[t]
  \centering
  \includegraphics[width=0.8\columnwidth]{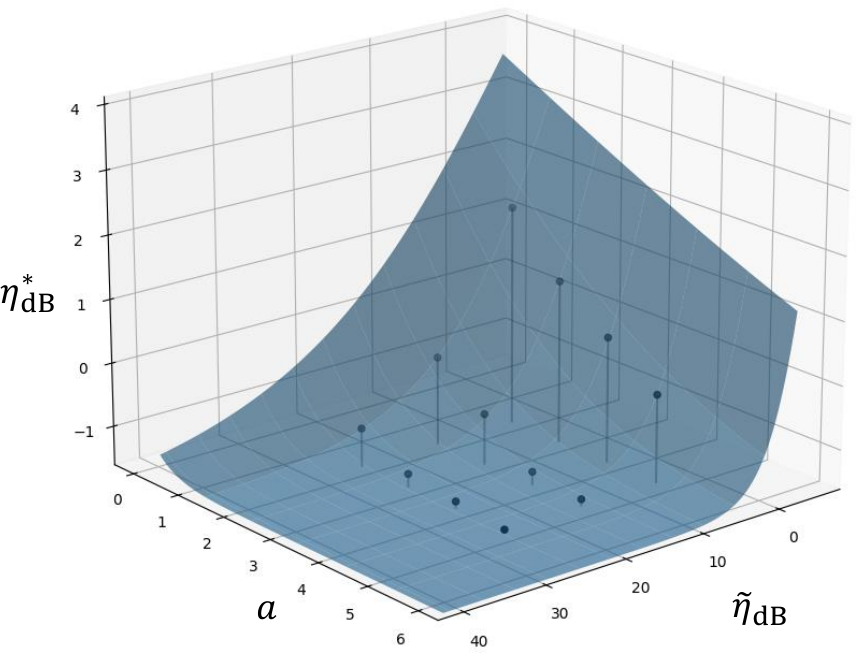}\\
  \caption{The logistic function in \eqref{eq:fit} well approximate the required UL SNR of DEEP-IoT under the setup where $K=36$, $N=144$, and $R=1/4$.}
\label{fig:FS3D2}
\end{figure}

To achieve a specified target PER at the AP, we define $\eta^*_{\text{dB}}$ as the required UL SNR of DEEP-IoT. Through comprehensive numerical evaluations, we find that $\eta^*_{\text{dB}}$ adheres to an empirical model closely resembling a logistic function:
\begin{equation}\label{eq:fit}
    \eta^*_{\text{dB}} = \frac{1}{ \exp\left(u_0\widetilde{\eta}_{\text{dB}}+u_1 a+u_2\widetilde{\eta}_{\text{dB}} a + u_3\right) + u_4} + u_5,
\end{equation}
where $\{u_0,u_1,u_2,u_3,u_4,u_5\}$ are constants.
It is important to highlight that an elevated DL SNR invariably benefits DEEP-IoT, leading to a reduction in the required UL SNR to meet the target PER. Therefore, $\forall a_{\ell}\geq 0$, we have
\begin{eqnarray*}
&&\hspace{-0.65cm}
\frac{\partial \eta^*_{\text{dB}}}{\partial \widetilde{\eta}_{\text{dB}}}\!=\!
-\!\left[\frac{1}{ \exp\!\left(u_0\widetilde{\eta}_{\text{dB}}\!+\!u_1 a_{\ell}\!+\!u_2\widetilde{\eta}_{\text{dB}} a_{\ell} \!+\! u_3\right) \!+\! u_4}\right]^2 \\
&&\hspace{0.3cm}
\exp\!\left(u_0\widetilde{\eta}_{\text{dB}}+u_1 a_{\ell}+u_2\widetilde{\eta}_{\text{dB}} a_{\ell} + u_3\right)(u_0+u_2 a_{\ell}) \leq 0.
\end{eqnarray*}
This suggest that $u_2>0$ since $u_0$ is a constant.

Within the parameters specified in Table \ref{tab:params}, Fig. \ref{fig:FS3D} corroborates the empirical model described by \eqref{eq:fit} as a viable approximation of the interplay among $\eta^*_{\text{dB}}$, $\widetilde{\eta}_{\text{dB}}$, and $a$. The constants for this model are determined as follows:
$u_0 = 0.08$
$u_1 = 0.5$,
$u_2 = 0.05$,
$u_3 = -2.65$,
$u_4 = 0.116$, and
$u_5 = -1.22$.
To further substantiate this approximation, a distinct system configuration featuring $K=36$, $N=144$, and a coding rate $R=1/4$ is examined. Targeting a PER of $10^{-4}$, Fig. \ref{fig:FS3D2} demonstrates that \eqref{eq:fit} accurately estimates the required UL SNR for DEEP-IoT under these conditions by adjusting the constants to
$u_0 = 0.073$
$u_1 = 0.4$,
$u_2 = 0.05$,
$u_3 = -1.92$,
$u_4 = 0.085$, and
$u_5 = -1.8$.
These findings validate the logistic function as a reliable method for predicting the necessary UL SNR in DEEP-IoT systems, illustrating its effectiveness across varying system settings.

\subsubsection{Average transmit power} 
Given \eqref{eq:fit}, we proceed to calculate the average transmit power required for DEEP-IoT. 

\begin{prop}\label{thm:prop1}
In Rayleigh fading channels, the average transmit power of DEEP-IoT is
\begin{equation}\label{eq:EP}
% \mathbb{E}P=P_{\max}\big[1-F_{P}(P_{\max})\big]+\int_{0}^{P_{\max}} p F'_{P}(p)dp.
\mathbb{E}P=\int_{0}^{\infty} \max(p,P_{\max}) F'_{P}(p)dp,
\end{equation}    
where the probability density function (PDF) of the transmit power $F'_{P}(p)$ is detailed in \eqref{eq:pdf_p}.
\end{prop}

\begin{NewProof}
To start with, we ignore the maximum power constraint $P_{\max}$ and derive the cumulative distribution function (CDF) of the transmit power $P$.
Substituting \eqref{eq:III2} and \eqref{eq:III5} into \eqref{eq:fit} yields
\begin{equation}
    10\lg \frac{\alpha|h|^2}{N_0}P = 
    \Big[ e^{10(u_0+u_2a)\lg\frac{\alpha|\widetilde{h}|^2}{N_0}\widetilde{P}_{\max}}e^{u_1a+u_3} +u_4 \Big]^{-1}\!\!+u_5.
\end{equation}

Since 
\begin{equation*}
e^{10(u_0+u_2a)\lg\frac{\alpha|\widetilde{h}|^2}{N_0}\widetilde{P}_{\max}} 
= \Big(\frac{\alpha|\widetilde{h}|^2}{N_0}\widetilde{P}_{\max}\Big)^{\frac{10}{\ln 10}(u_0+u_2a)},
\end{equation*}
we have
\begin{eqnarray}\label{eq:P}
&&\hspace{-0.5cm}P=\frac{N_0}{\alpha|h|^2}10^{\frac{u_5}{10}}
\exp\Bigg\{\frac{\ln 10}{10}
\bigg[\Big(\frac{\alpha|\widetilde{h}|^2}{N_0}\widetilde{P}_{\max}\Big)^{\frac{10}{\ln 10}(u_0+u_2a)} \nonumber\\
&&\hspace{3cm}
e^{u_1a+u_3} +u_4 \bigg]^{-1}
\Bigg\} \nonumber\\
&&\hspace{-0.1cm}
\triangleq
U_0 \frac{1}{|h|^2} \exp \Bigg\{ \frac{U_1}{U_2 \Big(\big|\widetilde{h}\big|^2\Big)^{U_3}+u_4} \Bigg\},
\end{eqnarray}
where
$U_0=\frac{N_0}{\alpha}10^{\frac{u_5}{10}}>0$,
$U_1=\frac{\ln 10}{10}>0$,
$U_2=\Big(\frac{\alpha}{N_0}\widetilde{P}_{\max}\Big)^{\frac{10}{\ln 10}(u_0+u_2a)}
e^{u_1a+u_3}>0$, and
$U_3=\frac{10}{\ln 10}(u_0+u_2a)>0$.

Let $V\triangleq\frac{1}{|h|^2}$ and $S\triangleq\exp \Big\{ \frac{U_1}{U_2 \big(|\widetilde{h}|^2\big)^{U_3}+u_4} \Big\}$.
Since $|h|^2\sim \text{Exp}\left(\frac{1}{\sigma^2}\right)$, it can be shown that the CDF and PDF of $V$ are 
\begin{equation*}
F_V(v)=\exp\Big(-\frac{1}{\sigma^2 v}\Big),~~
F'_V(v) = \frac{1}{\sigma^2 v^2} \exp\Big(-\frac{1}{\sigma^2 v}\Big),
\end{equation*}
respectively. 
On the other hand, since $|\widetilde{h}|^2\sim \text{Exp}\left(\frac{1}{\widetilde{\sigma}^2}\right)$, the CDF of $S$ can be written as
\begin{eqnarray*}
F_S(s) \hspace{-0.2cm}&=&\hspace{-0.2cm} 1 -
\Pr\Bigg(|\widetilde{h}|^2\leq\Big(\frac{U_1}{U_2\ln s}-\frac{u_4}{U_2}\Big)^{\frac{1}{U_3}}\Bigg) \\
\hspace{-0.2cm}&=&\hspace{-0.2cm}
\begin{cases}
    \exp \bigg\{ -\frac{1}{\widetilde{\sigma}^2} \Big(\frac{U_1}{U_2\ln s}-\frac{u_4}{U_2}\Big)^{\frac{1}{U_3}} \bigg\}, & \hspace{-0.2cm}\text{if}~1\leq s \leq e^{\frac{U_1}{u_4}},  \\
    1, & \hspace{-0.2cm}\text{if}~s > e^{\frac{U_1}{u_4}}.
\end{cases}
\end{eqnarray*}
Therefore, the PDF of $S$ is
\begin{eqnarray*}
F'_S(s)
\hspace{-0.2cm}&=&\hspace{-0.2cm}
\frac{U_1}{\widetilde{\sigma}^2U_3U_2^{\frac{1}{U_3}}}
\frac{ \Big(U_1-u_4\ln s\Big)^{\frac{1}{U_3}-1} }{s(\ln s)^{\frac{1}{U_3}+1}}
\exp \bigg\{ -\frac{1}{\widetilde{\sigma}^2} \\
&&\hspace{-0.2cm}
\Big(\frac{U_1}{U_2\ln s}-\frac{u_4}{U_2}\Big)^{\frac{1}{U_3}} \bigg\},
~~~~ 1\leq s \leq e^{\frac{U_1}{u_4}}.
\end{eqnarray*}

Given $F'_V(v)$ and $F'_S(s)$, the CDF of the transmit power $P$ of the IoT device can be written as
\begin{eqnarray*}
F_{P}(p)
\hspace{-0.2cm}&=&\hspace{-0.2cm}
\iint_{vs\leq \frac{p}{U_0}} F'_V(v) F'_S(s) dv ds \\
\hspace{-0.2cm}&=&\hspace{-0.2cm}
\int_{1}^{e^{\frac{U_1}{u_4}}} F_V\left(\frac{p}{U_0 s}\right) F'_S(s) ds \\
\hspace{-0.2cm}&\overset{(a)}{=}&\hspace{-0.2cm}
\frac{U_1}{\widetilde{\sigma}^2U_3U_2^{\frac{1}{U_3}}}
\int_{1}^{e^{\frac{U_1}{u_4}}}\frac{1}{r^2}
\Big(\frac{U_1}{r}-u_4\Big)^{\frac{1}{U_3}-1}
\exp \bigg\{ -\frac{1}{\widetilde{\sigma}^2} \\
&&\hspace{-0.2cm}
\Big(\frac{U_1}{U_2r}-\frac{u_4}{U_2}\Big)^{\frac{1}{U_3}}-\frac{U_0e^r}{\sigma^2 p} \bigg\} dr,
\end{eqnarray*}
where (a) follows by defining $r\triangleq \ln y $. Correspondingly, the PDF of $P$ can be derived as:
\begin{eqnarray}\label{eq:pdf_p}
F'_{P}(p) \hspace{-0.2cm}&=&\hspace{-0.2cm}
\frac{U_0U_1}{p^2\sigma^2\widetilde{\sigma}^2U_3U_2^{\frac{1}{U_3}}}
\int_{1}^{e^{\frac{U_1}{u_4}}}\frac{1}{r^2}
\Big(\frac{U_1}{r}-u_4\Big)^{\frac{1}{U_3}-1}
\nonumber\\
&&\hspace{-1cm}
\exp \bigg\{ -\frac{1}{\widetilde{\sigma}^2}\Big(\frac{U_1}{U_2r}-\frac{u_4}{U_2}\Big)^{\frac{1}{U_3}}-\frac{U_0e^r}{\sigma^2 p} + r \bigg\} dr.
\end{eqnarray}

Finally, combining the distribution of $P$ with the maximum power constraint $P\leq P_{\max}$, we obtain the average transmit power of DEEP-IoT in \eqref{eq:EP}.
\end{NewProof}

\begin{figure}[t]
  \centering
  \includegraphics[width=0.8\columnwidth]{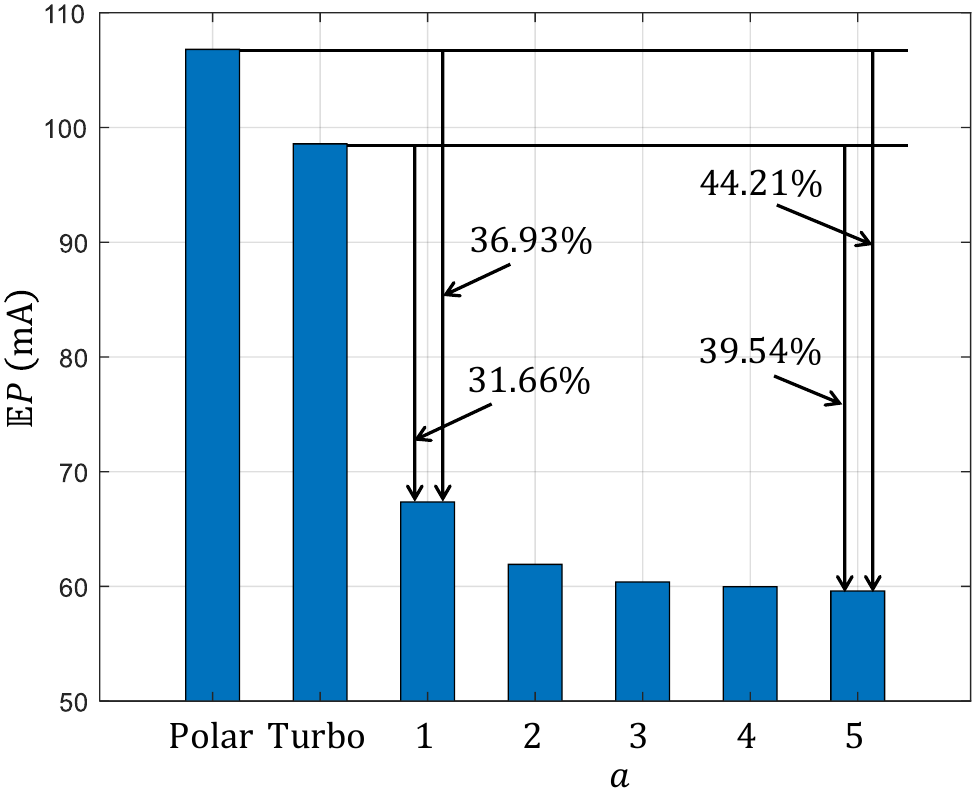}\\
  \caption{A numerical evaluation of the average transmit power of DEEP-IoT (with different feedback subcarriers $a$) benchmarked against conventional IoT system with Polar and Turbo codes.}
\label{fig:bar}
\end{figure}

In line with Proposition \ref{thm:prop1}, we conduct a numerical evaluation of DEEP-IoT's average transmit power, as depicted in Fig.~\ref{fig:bar}. This comparison highlights DEEP-IoT's remarkable capability to lower the average transmit power substantially when contrasted with conventional IoT systems utilizing Polar and Turbo codes. Notably, with the deployment of a single feedback subcarrier ($a=1$), DEEP-IoT secures reductions in average transmit power of $36.93\%$ and $31.66\%$ relative to systems employing Polar and Turbo codes, respectively. The efficiency gain is even more significant as the number of allocated feedback subcarriers increases: at $a=5$, the reduction in average transmit power escalates to $44.21\%$ and $39.54\%$, respectively. Consequently, this pronounced decrease in power consumption effectively prolongs the IoT device's operational lifespan to approximately $\frac{1}{1-44.21\%}\approx 1.80$ and $\frac{1}{1-39.54\%}\approx 1.65$ times, respectively. Such findings underscore DEEP-IoT's profound contributions towards boosting energy efficiency and extending the longevity of devices across IoT networks, marking a substantial leap forward in sustainable IoT communication practices.

\section{DEEP-IoT MAC Layer}\label{sec:V}
Section \ref{sec:IV} details the DEEP-IoT physical layer and its effectiveness in reducing the energy consumption of individual IoT devices. This section shifts focus to the MAC layer of DEEP-IoT, aiming to assess the potential enhancements in lifespan for an IoT cell comprising multiple devices. A key differentiator of the DEEP-IoT system at the MAC layer, when compared to traditional communication systems, is the incorporation of feedback as a new data type. This integration necessitates a reevaluation of resource allocation strategies, particularly the need for managing both UL data and associated DL feedback within a dual-channel framework. 

As outlined in Section \ref{sec:III}, this paper adopts the HD-FDD mode to tackle the challenge of dual channel access in the DEEP-IoT MAC layer. Similar to NB-IoT, the approach involves using narrow-band SC-FDMA for the UL data transmission from IoT devices and OFDMA for the AP’s UL feedback and data transmission. This design ensures that the UL data and the corresponding DL feedback are interwoven in the time domain and distinctly allocated across various subcarriers in the frequency domain. Such a configuration in the MAC protocol effectively simplifies the process, enabling a concentrated focus on the resource allocation specifically for the downlink feedback channel.

\subsection{Feedback channel allocation}
We consider a joint sensing task of IoT devices. Due to the distributed deployment of IoT devices, they experience diverse channel conditions with the AP. Over time, the communication energy consumption of these IoT devices varies, and the depletion of energy in any single device leads to the failure of the joint sensing task.

In legacy communication systems, devices under poorer channel conditions consume more energy for data transmission to achieve the desired decoding performance. Under such circumstances, the life expectancy of the sensing IoT network is capped by the device with the worst channel condition. 
DEEP-IoT offers an innovative approach to this dilemma: by strategically allocating more feedback to devices in poorer channel conditions and with lower energy reserves, we can boost their coding efficiency and reduce their energy consumption, ultimately extending the lifespan of the entire IoT cell. In the following, we shall formulate the feedback channel allocation problem and examine the potential gain of DEEP-IoT.

We start by introducing some definitions:
\begin{itemize}
    \item The overall lifespan of the IoT cell is $T$ cycles, and the initial energy reserve of the $\ell$-th IoT device is $\rho_{\ell}$.
    \item At the beginning of the $t$-th cycle, the remaining energy reserve of the $\ell$-th IoT device is $\text{E}^{(t)}_{\ell}$.
    \item In the $t$-th cycle, the energy consumed by the $\ell$-th IoT device for transmitting, receiving, and sleep are denoted by $\text{ET}^{(t)}_{\ell}$, $\text{ER}^{(t)}_{\ell}$, and $\text{ES}^{(t)}_{\ell}$, respectively.
    \item The receive power and sleep power of an IoT device are denoted by $P_r$ and $P_s$, respectively.
    \item The duration of an OFDM symbol is $T_{\text{OFDM}}$. %$300$us.
\end{itemize}

Recall that the number of feedback subcarriers allocated to the $\ell$-th IoT device in the $t$-th cycle is $a^{(t)}_{\ell}\in\mathbb{N}^+$, the feedback channel allocation problem can be formulated as follows:

\begin{subequations}\label{eq:P1}
\begin{align}
&\hspace{-0.2cm}\max_{\{a^{(t)}_{\ell}\}} \mathbb{E}
\left[T:\text{E}^{(T)}_{\ell}\geq 0,\forall \ell, \left.\min_{\ell} \text{E}^{(T+1)}_{\ell}< 0 \right\vert \text{E}^{(1)}_{\ell}=\rho_{\ell} \right]
\label{eq:P1a}\\
& { ~\text {s.t.}~} 
\text{E}^{(t+1)}_{\ell}=\text{E}^{(t)}_{\ell}-\text{ET}^{(t)}_{\ell}-\text{ER}^{(t)}_{\ell}-\text{ES}^{(t)}_{\ell},
\label{eq:P1b}\\
&\hphantom {~\text {s.t.}~} 
\text{ET}^{(t)}_{\ell}=P^{(t)}_{\ell}T_{\text{OFDM}}QG,
\label{eq:P1c}\\
&\hphantom {~\text {s.t.}~}
\text{ER}^{(t)}_{\ell}=\left[1-\delta(a^{(t)}_{\ell}=0)\right]P_rT_{\text{OFDM}}Q(G-1),
\label{eq:P1d}\\
&
\hphantom {~\text {s.t.}~} 
\text{ES}^{(t)}_{\ell}=\delta(a^{(t)}_{\ell}=0)P_sT_{\text{OFDM}}Q(G-1),
\label{eq:P1e}\\
&\hphantom {~\text {s.t.}~}
P^{(t)}_{\ell}=\min\left\{P_{\max},\eta^*\!\left(\widetilde{h}^{(t)}_{\ell},a^{(t)}_{\ell}\right)\frac{N_0}{\alpha_{\ell}|h^{(t)}_{\ell}|^2}\right\},
\label{eq:P1f}\\
&\hphantom {~\text {s.t.}~}
\sum_{\ell=1}^{L}a^{(t)}_{\ell}\leq M,~\forall t.
\label{eq:P1g}
\end{align}
\end{subequations}

% \begin{eqnarray}\label{eq:P1}
% &&\hspace{-1.2cm}
% \max_{\{a^{(t)}_{\ell}\}}~\mathbb{E}
% \left[T:\text{E}^{(T)}_{\ell}\geq 0,\forall \ell, \left.\min_{\ell} \text{E}^{(T+1)}_{\ell}< 0 \right\vert \text{E}^{(1)}_{\ell}=\rho_{\ell} \right] \\
% &&\hspace{-0.9cm}
% \text {s.t.} ~~~~
% \text{E}^{(t+1)}_{\ell}=\text{E}^{(t)}_{\ell}-\text{ET}^{(t)}_{\ell}-\text{ER}^{(t)}_{\ell}-\text{ES}^{(t)}_{\ell},
% \nonumber\\
% &&
% \text{ET}^{(t)}_{\ell}=P^{(t)}_{\ell}T_{\text{OFDM}}QG, 
% \nonumber\\
% &&
% \text{ER}^{(t)}_{\ell}=\left[1-\delta(a^{(t)}_{\ell}=0)\right]P_rT_{\text{OFDM}}Q(G-1),
% \nonumber\\
% &&
% \text{ES}^{(t)}_{\ell}=\delta(a^{(t)}_{\ell}=0)P_sT_{\text{OFDM}}Q(G-1),
% \nonumber\\
% &&
% P^{(t)}_{\ell}=\min\left\{P_{\max},10^{\frac{1}{10}\eta^*_{\text{dB}}\left(\widetilde{h}^{(t)}_{\ell},a^{(t)}_{\ell}\right)}\frac{N_0}{\alpha_{\ell}|h^{(t)}_{\ell}|^2}\right\},
% \nonumber\\
% &&
% \sum_{\ell=1}^{L}a^{(t)}_{\ell}\leq M,~\forall t.
% \nonumber
% \end{eqnarray}

This formulation underscores our primary objective: to extend the operational lifespan of the IoT network, i.e., $T$ in \eqref{eq:P1a}, by judiciously distributing subcarriers among IoT devices.
Eq. \eqref{eq:P1b} to \eqref{eq:P1e} quantify the energy consumption of each IoT device within a given cycle, accounting for activities such as transmission, receiving, and periods of inactivity.
In each cycle, the transmit power of an IoT device is configured based on the feedback channel quality and the number of subcarriers assigned to the device, as specified in \eqref{eq:P1f}. Additionally, \eqref{eq:P1g} delineates the constraints on the total allocation of feedback subcarriers.

\subsection{DP with function approximation}
Given that the subcarriers are allocated across successive cycles, \eqref{eq:P1} is essentially a sequential decision-making problem. Consequently, it is feasible to model \eqref{eq:P1} as a Markov Decision Process (MDP) and approach its resolution through dynamic programming (DP).
In the beginning of the $t$-th cycle, we characterize the state of the IoT cell as $\bm{s}^{(t)} \triangleq \left\{\left[\text{E}^{(t)}_{\ell},h^{(t)}_{\ell},\widetilde{h}^{(t)}_{\ell}\right]: \forall \ell  \right\}$. In particular, a state $\bm{s}^{(t)}$ is designated as terminal if and only if $\min_{\ell} \text{E}^{(t)}_{\ell}< 0$. The collection of all terminal states is denoted by $\mathcal{S}_{\text{ter}}$.
The AP implements a subcarrier allocation policy $\pi$ that dictates the allocation actions, i.e., the subcarrier distributions for the $t$-th cycle $\bm{a}^{(t)}\triangleq\left\{a^{(t)}_{\ell}: \forall \ell\right\}$, based on the current state. That is, $\bm{a}^{(t)}=\pi\big(\bm{s}^{(t)}\big)$. The cardinality of the action space $\mathcal{A}$ is $|\mathcal{A}|=\sum_{i=0}^{M}\tbinom{L+i-1}{L-1}$.

Upon executing action $a^{(t)}_{\ell}$, each IoT device tunes its transmit power $P^{(t)}_\ell$ based on the UL and DL channel gains, $h^{(t)}_{\ell}$ and $\widetilde{h}^{(t)}_{\ell}$, and communicates with the AP with the DEEP-IoT physical layer specified in Section \ref{sec:IV}.
Transitioning to the start of the $(t+1)$-th cycle, the system evolves into a new state, $\bm{s}^{(t+1)}$, in accordance with the dynamics outlined in \eqref{eq:P1}. Given that $\text{E}^{(t)}_{\ell} < \text{E}^{(t+1)}_{\ell}$, $\forall\ell$, it is inevitable for the system state to eventually reach a terminal state, positioning the MDP within the scope of a finite-horizon problem. A cycle yields an instantaneous reward of $r^{(t)}=1$, provided that the system state $\bm{s}^{(t)}$ is not a terminal state. 

Overall, the objective of the MDP is to maximize the cumulative reward, $\sum_{t=1}^{T} r^{(t)}=T$, across an episode. Here, $T$ denotes the cycle at which the system first encounters a terminal state within an episode, aligning this optimization target with the goal of the original feedback channel allocation problem in \eqref{eq:P1}.
The optimal solution to the finite horizon MDP, and hence to \eqref{eq:P1}, can be achieved through DP as follows:
\begin{equation}\label{eq:via}
    q_*(\bm{s},\bm{a})=\delta(\bm{s}\notin\mathcal{S}_{\text{ter}}) + \mathbb{E}_{\bm{s}^\prime}\left[ \max_{\bm{a}^\prime} q_*(\bm{s}',\bm{a}')\right],
\end{equation}
where $q_*(\bm{s},\bm{a})$ represents the optimal action-value function. 

The continuous nature of the state space $\bm{s}$, however, renders the direct application of the value iteration algorithm (VIA) impractical due to the necessity of state space quantization, which is computationally prohibitive. To navigate this challenge, we adopt value function approximation, i.e., estimate the optimal action-value function through a function approximation, thereby facilitating efficient computation and circumventing the direct quantization of the continuous state space.
Specifically, we approximate the optimal action-value function by:
\begin{equation}\label{eq:Qapprox}
    q_*(\bm{s},\bm{a})\approx\Phi_{\varphi}(\bm{s},\bm{a}),
\end{equation}
where $\Phi_{\varphi}$ denotes a DNN parameterized by a vector of weights $\varphi$. The discovery of an appropriate vector of DNN weights $\varphi$ that fulfills \eqref{eq:Qapprox} paves the way to determining the optimal policy $\pi_*$.
To this end, for a pair of state $\bm{s}= \Big\{\big[\text{E}_{\ell},h_{\ell},\widetilde{h}_{\ell}\big]: \forall \ell  \Big\}$ and action $\bm{a}$, we introduce a quadratic loss function $\mathcal{L}(\bm{s},\bm{a};\varphi)$ to facilitate the training of $\varphi$ through gradient descent:
\begin{eqnarray}
&&\hspace{-0.5cm} \bm{s}^\prime = \left\{\left[\text{E}_{\ell}-\text{ET}_{\ell}-\text{ER}_{\ell}-\text{ES}_{\ell},h^\prime_{\ell},\widetilde{h}^\prime_{\ell}\right]: \forall \ell  \right\},
\label{eq:DPsnew}\\
&&\hspace{-0.5cm} \mathcal{L}(\bm{s},\bm{a};\varphi)= \Big[\Phi_{\varphi}(\bm{s},\bm{a})- \delta(\bm{s}\notin\mathcal{S}_{\text{ter}}) 
\label{eq:DPL2}\\
&&\hspace{0.1cm} -\int_{\{h^\prime_{\ell},\widetilde{h}^\prime_{\ell}\}}
     \max_{\bm{a}^\prime} \Phi_{\varphi}(\bm{s}^\prime,\bm{a}^\prime)\Pr\big(\{h^\prime_{\ell},\widetilde{h}^\prime_{\ell}\}\big)
    d\{h^\prime_{\ell},\widetilde{h}^\prime_{\ell}\} \Big]^2,
\nonumber\\
&&\hspace{-0.5cm}     \varphi \leftarrow \varphi - \epsilon\nabla_{\varphi}\mathcal{L}(\bm{s},\bm{a};\varphi),\label{eq:DPupdate}
\end{eqnarray}
where $\epsilon$ is the step size. Following the practice of reinforcement learning (RL) \cite{atari}, we approximate the $\Phi_{\varphi}$ within the $\max$ operator in \eqref{eq:DPL2} by another DNN. This dual network architecture helps to mitigate the correlation between successive state-action pairs' approximated $q$ values, enhancing the stability and efficiency of the learning process.

\subsection{Index table-based policy gradient}
Employing function approximation to directly estimate the optimal action-value function in DP obviates the need for discretizing the state space, thereby avoiding the quantization errors. Furthermore, it capitalizes on the known dynamics of the system to forge a more robust solution.

Nonetheless, integrating DP with function approximation presents its own set of hurdles. A particularly daunting challenge is the requirement for exhaustive integral calculations across the UL and DL channel realizations for all IoT devices, as delineated in \eqref{eq:DPL2}. These integrations call for significant computational effort, especially when the channel gain distributions are widespread and varied. This complexity stems from the inherent nature of full-width backup in DP -- assessing the value of any state-action pair involves accounting for every conceivable subsequent state-action pairs. As the dimensionality of the problem escalates, the practicality of employing DP in conjunction with function approximation begins to wane due to these computational demands. To tackle this, we resort to sampling-based RL approaches.

% Sampling-based RL approaches, such as Monte Carlo methods or Temporal Difference learning, can significantly reduce the computational overhead by selectively sampling state-action pairs rather than exhaustively computing the integral across all possible outcomes.

For any subcarrier allocation policy $\pi$, we can sample a history of state transitions, known as a {\it trajectory}: $\tau = \{\bm{s}^{(1)},\bm{a}^{(1)},r^{(1)},\bm{s}^{(2)},\bm{a}^{(2)},r^{(2)},...,\bm{s}^{(T)},\bm{a}^{(T)},r^{(T)}\}$.
The objective is to refine the policy $\pi$ to optimize the average sampled rewards. This approach is a cornerstone of policy gradient algorithms \cite{PG}. In this context, we can directly model the policy $\pi$ by a DNN $\Psi_{\psi}$, characterized by a vector of parameters $\psi$. The policy is then iteratively updated to maximize the average total reward:
\begin{equation}\label{eq:PGupdate}
\psi \leftarrow \psi + \epsilon\nabla_{\psi}\mathbb{E}_{\tau\sim\pi_{\psi}}\big[T(\tau)\big],
\end{equation}
where $T(\tau)$ denotes the lifespan of the IoT cell in a trajectory $\tau$. It is easy to show that
\begin{equation*}
\nabla_{\psi}\mathbb{E}_{\tau}\big[T(\tau)\big]
=\mathbb{E}_{\tau}\Bigg[\sum_{t=1}^{T(\tau)}\nabla_{\psi}\ln \pi_{\psi}\big( a^{(t)}(\tau)|s^{(t)}(\tau) \big)T(\tau)\Bigg],
\end{equation*}
where
\begin{equation*}
\Pr(\tau)\!=\!\!\sum_{t=1}^{T(\tau)}\!\pi_{\psi}\Big( a^{(t)}\!(\tau)|s^{(t)}\!(\tau) \Big)\!\Pr\Big(\{h^{(t+1)}_{\ell}\!(\tau),\widetilde{h}^{(t+1)}_{\ell}\!(\tau)\}\Big).
\end{equation*}

\begin{rem}
While this sampling-based Monte Carlo approach diverges from DP by not using the known system state transitions, it can implicitly learn about the system's dynamics through extensive and systematic sampling of state transitions, allowing for a more effective means of policy optimization to maximize rewards. 
Furthermore, the introduction of policy parameterization opens the door to new and low-complexity solutions, such as the index table-based scheme proposed below.
\end{rem}

In DP with function approximation, the DNN $\Phi_{\phi}$ is tasked with approximating the $q$ values corresponding to specific state-action pairs: the input dimensionality is $3L + \sum_{i=0}^{M}\tbinom{L+i-1}{L-1}$, signifying the state and possible actions, while the output dimension is a singular value, representing the estimated $q$ value. Conversely, within the framework of standard policy gradient methods, the DNN $\Phi_{\varphi}$ is employed to parameterize the policy itself. The input dimensionality of this configuration is $3L$, while the output dimensionality is $\sum_{i=0}^{M}\tbinom{L+i-1}{L-1}$, corresponding to the probability distribution over the action space.
As can be seen, despite policy gradient methods circumventing the need to compute integrals numerically, the inherent training complexity remains directly proportional to the number of IoT devices $L$. This relationship poses substantial challenges in scenarios where $L$ is large, complicating the training process due to the exponential increase in computational demands.

Addressing this challenge, this paper puts forth a new, index table-based, low-complexity policy gradient method to solve \eqref{eq:P1} by leveraging insights from multi-armed bandit allocation indices \cite{gittins}. The cornerstone of this approach is its emphasis on dimensionality reduction. By decoupling the $L$-dimensional MDP into $M$ one-dimensional MDPs, we achieve a significant reduction in complexity. Specifically, this method ensures that the computational complexity grows linearly with $L$, and crucially, the training complexity remains invariant regardless of the number of IoT devices $L$. 

Let us consider a single IoT device and examine how varying the number of feedback subcarriers influences its state transitions. Defining $\bm{s}^{(t)}_{\ell} \triangleq \big[\text{E}^{(t)}_{\ell},h^{(t)}_{\ell},\widetilde{h}^{(t)}_{\ell}\big]$, the transition of $\bm{s}^{(t)}_{\ell}$ is a one-dimensional MDP. Therefore, the optimal action-value function can be written as 
\begin{eqnarray}\label{eq:via1}
q_*(\bm{s}_{\ell},a_{\ell})
\hspace{-0.2cm}&=&\hspace{-0.2cm}
\frac{1}{L}\delta(\text{E}_{\ell}>0) + \mathbb{E}_{\bm{s}^\prime_{\ell}}\left[ \max_{a^\prime_{\ell}} q_*(\bm{s}'_{\ell},a'_{\ell})\right], \nonumber\\
\hspace{-0.2cm}&\triangleq&\hspace{-0.2cm}
r_{\ell} + \mathbb{E}_{\bm{s}^\prime_{\ell}}\big[ v_*(\bm{s}^\prime_{\ell})\big],
\end{eqnarray}
where the instantaneous reward of the IoT device is defined as $r_{\ell}=\frac{1}{L}\delta(\text{E}_{\ell}>0)$ because each device aims to maximize the operational longevity by ensuring its energy level remains above zero; $v_*(\bm{s}_{\ell})$ is the value function of the state $\bm{s}^\prime_{\ell}$, reflecting the long-term accumulated reward that can be reaped from being in state $\bm{s}^\prime_{\ell}$. In the scenario of a single-device MDP, the state transitions are solely influenced by the actions taken for that particular device $a_{\ell}$, but not $a_{\ell^{\prime}}$, $\ell^{\prime}\neq\ell$.

\begin{lem}[monotonicity of the value function]\label{thm:lem1}
    Consider any two states of the single-device MDP that share the same channel realizations: $\bm{s}_{\ell} = \left[\text{E}_{\ell},h_{\ell},\widetilde{h}_{\ell}\right]$ and $\bm{s}^\prime_{\ell} = \left[\text{E}^\prime_{\ell},h_{\ell},\widetilde{h}_{\ell}\right]$. If $\text{E}_{\ell}\leq \text{E}^\prime_{\ell}$,
    \begin{equation}
        v_*(\bm{s}_{\ell})\leq v_*(\bm{s}^\prime_{\ell}).
    \end{equation}
    That is, $v_*(\bm{s}_{\ell})$ is monotonically non-decreasing in $\text{E}_{\ell}$ for a given pair of $h_{\ell}$ and $\widetilde{h}_{\ell}$.
\end{lem}

\begin{NewProof}
See Appendix \ref{sec:AppA}.
\end{NewProof}

To evaluate the effectiveness of an action $a_{\ell}$ when in a state $\bm{s}_{\ell}$, we set $a_{\ell}=0$ as a reference point, and define an index for each action $a_{\ell}>0$. The index will provide a quantitative measure to assess each action's relative benefit compared to the baseline.

\begin{defi}[index of an action]
In any state $\bm{s}_{\ell}$, we define an index for each action $a_{\ell}>0$:
\begin{equation}\label{eq:defidx}
\mu(\bm{s}_{\ell},a_{\ell}) \triangleq q_*(\bm{s}_{\ell},a_{\ell}) - q_*(\bm{s}_{\ell},0). 
\end{equation}
\end{defi}

Intuitively, the index of an action $\mu(a_{\ell})$ can be understood as an allocating benefits, representing the benefits that can be derived from allocating $a_{\ell}$ subcarriers to the IoT device.
% In essence, this index quantifies how beneficial it is for the AP to expend resources to the device. 
Theorem~\ref{thm:1} below summarizes the main properties of the indexes.

\begin{thm}[properties of the indexes]\label{thm:1}
Consider a single-device MDP and a state $\bm{s}_{\ell} = \big[\text{E}_{\ell},h_{\ell},\widetilde{h}_{\ell}\big]$.
\begin{itemize}[leftmargin=0.38cm]
    \item Monotonicity. The indexes $\mu(\bm{s}_{\ell},a_{\ell})$, $\forall a_{\ell}> 0$, are monotonic: when the DL channel SNR $\widetilde{\eta}_{\text{dB}}\geq -\frac{u_1}{u_2}$, $\mu(\bm{s}_{\ell},a_{\ell})$ is monotonically non-decreasing in $a_{\ell}$; otherwise, when $\widetilde{\eta}_{\text{dB}}< -\frac{u_1}{u_2}$, $\mu(\bm{s}_{\ell},a_{\ell})$ is monotonically non-increasing in $a_{\ell}$.
    \item Sign. The index $\mu(\bm{s}_{\ell},a_{\ell})\leq 0$ if and only if
\begin{equation} \eta^*\!\left(\widetilde{h}_{\ell},a_{\ell}\right)
\geq
\min\Big\{\eta^{\text{max}}_{\ell},\eta^*_{\ell}(0) \Big\} -\frac{G-1}{G}(P_r\!-\! P_s)\allowbreak\frac{\alpha_{\ell}|h_{\ell}|^2}{N_0},
\end{equation}
where $\eta^{\text{max}}_{\ell}\triangleq \frac{\alpha_{\ell}|h_{\ell}|^2}{N_0}P_{\max}$ denotes the maximum UL received SNR at the AP when the IoT device transmits at its maximum power $P_{\max}$; $\eta_{\ell}^*(0)$ denotes the minimum UL SNR required for conventional forward channel codes to achieve a target PER.\footnote{Given a specific UL channel gain $h_{\ell}$, both $\eta_{\ell}^{\text{max}}$ and $\eta_{\ell}^*(0)$ are constants. The value of $\eta_{\ell}^*(0)$ is determined by the choice of the conventional forward channel code, e.g., Polar and Turbo codes.}
\end{itemize}
\end{thm}

\begin{NewProof}
We first prove the monotonicity of the indexes.
From the definition of the index in \eqref{eq:defidx}, we have
\begin{eqnarray}
&&\hspace{-0.5cm}\frac{\partial \mu(\bm{s}_{\ell},a_{\ell})}{\partial a_{\ell}} 
= \frac{\partial q_*(\bm{s}_{\ell},a_{\ell})}{\partial a_{\ell}} \\
&&\hspace{-0.5cm}
= \int_{h^\prime_{\ell},\widetilde{h}^\prime_{\ell}}\frac{\partial v_*\left(\bm{s}^\prime_{\ell}=\left[\text{E}^\prime_{\ell},h^\prime_{\ell},\widetilde{h}^\prime_{\ell}\right]\right)}{\partial a_{\ell}}\Pr(h^\prime_{\ell},\widetilde{h}^\prime_{\ell})d h^\prime_{\ell}d\widetilde{h}^\prime_{\ell} \nonumber\\
&&\hspace{-0.5cm}
=\mathbb{E}_{h^\prime_{\ell},\widetilde{h}^\prime_{\ell}}\left[\frac{\partial v_*\left(\bm{s}^\prime_{\ell}\right)}{\partial \text{E}^\prime_{\ell}}\frac{\partial \text{E}^\prime_{\ell}}{\partial a_{\ell}} \right].\nonumber
\end{eqnarray}

From Lemma \ref{thm:lem1}, we have $\frac{\partial v_*\left(\bm{s}^\prime_{\ell}\right)}{\partial \text{E}^\prime_{\ell}}\geq 0$. Furthermore,
\begin{eqnarray*}
&&\hspace{-0.5cm}
\frac{\partial \text{E}^\prime_{\ell}}{\partial a_{\ell}}=
-\frac{\partial \text{ET}^\prime_{\ell}}{\partial a_{\ell}} \\
&&\hspace{-0.5cm}
=
-\frac{\partial}{\partial a_{\ell}} \min\left\{P_{\max},10^{\frac{1}{10}\eta^*_{\text{dB}}\left(\widetilde{h}_{\ell},a_{\ell}\right)}\frac{N_0}{\alpha_{\ell}|h_{\ell}|^2}\right\}T_{\text{OFDM}}QG,
\end{eqnarray*}
where
\begin{eqnarray*}
&&\hspace{-0.65cm}
\frac{\partial \eta^*_{\text{dB}}\!\left(\widetilde{h}_{\ell},a_{\ell}\right)}{\partial a_{\ell}}\!=\!
-\!\left[\frac{1}{ \exp\!\left(u_0\widetilde{\eta}_{\text{dB}}\!+\!u_1 a_{\ell}\!+\!u_2\widetilde{\eta}_{\text{dB}} a_{\ell} \!+\! u_3\right) \!+\! u_4}\right]^2 \\
&&\hspace{0.7cm}
\exp\!\left(u_0\widetilde{\eta}_{\text{dB}}+u_1 a_{\ell}+u_2\widetilde{\eta}_{\text{dB}} a_{\ell} + u_3\right)(u_1+u_2\widetilde{\eta}_{\text{dB}}).
\end{eqnarray*}

Therefore, the sign of $\frac{\partial \text{E}^\prime_{\ell}}{\partial a_{\ell}}$ and $\frac{\partial \mu(\bm{s}_{\ell},a_{\ell})}{\partial a_{\ell}}$  depends on $u_1+u_2\widetilde{\eta}_{\text{dB}}$. 
Since $u_2>0$, when $\widetilde{\eta}_{\text{dB}}\geq-\frac{u_1}{u_2}$, $\frac{\partial \mu(\bm{s}_{\ell},a_{\ell})}{\partial a_{\ell}}\geq 0$; when $\widetilde{\eta}_{\text{dB}}<-\frac{u_1}{u_2}$, $\frac{\partial \mu(\bm{s}_{\ell},a_{\ell})}{\partial a_{\ell}}\leq 0$. The monotonicity of the indexes is proved.

Next, we prove the sign of the indexes.
The index of an action $a_{\ell}>0$ can be further written as:
\begin{eqnarray}\label{eq:APPC1}
&&\hspace{-0.5cm}
\mu(\bm{s}_{\ell},a_{\ell}) = q_*(\bm{s}_{\ell},a_{\ell}) - q_*(\bm{s}_{\ell},0) \\
&&\hspace{-0.5cm}
=\mathbb{E}_{h^\prime_{\ell},\widetilde{h}^\prime_{\ell}}\left[
v_*\left(\left[\text{E}^\prime_{\ell}(a_{\ell}),h^\prime_{\ell},\widetilde{h}^\prime_{\ell}\right]\right)-v_*\left(\left[\text{E}^\prime_{\ell}(0),h^\prime_{\ell},\widetilde{h}^\prime_{\ell}\right]\right)
\right],\nonumber
\end{eqnarray}
where
\begin{eqnarray*}
\text{E}^\prime_{\ell}(a_{\ell})
\hspace{-0.2cm}&=&\hspace{-0.2cm}
\text{E}_{\ell}-\text{ET}_{\ell}(h_{\ell},\widetilde{h}_{\ell},a_{\ell}) - \text{ER}_{\ell},
\\
\text{E}^\prime_{\ell}(0)
\hspace{-0.2cm}&=&\hspace{-0.2cm}
\text{E}_{\ell}-\text{ET}_{\ell}(h_{\ell},\widetilde{h}_{\ell},0) - \text{ES}_{\ell}.
\end{eqnarray*}

Since $v_*(\bm{s}_{\ell})$ is monotonic in $\text{E}_{\ell}$, the sign of $\mu(\bm{s}_{\ell},a_{\ell})$ depends on the sign of $\text{E}^\prime_{\ell}(a_{\ell})-\text{E}^\prime_{\ell}(0)$.
\begin{eqnarray}\label{eq:APPC2}
&&\hspace{-0.65cm}
\text{E}^\prime_{\ell}(a_{\ell})-\text{E}^\prime_{\ell}(0)
\\
&&\hspace{-0.65cm}
=\text{ET}_{\ell}(h_{\ell},\widetilde{h}_{\ell},0)-\text{ET}_{\ell}(h_{\ell},\widetilde{h}_{\ell},a_{\ell})+\text{ES}_{\ell}-\text{ER}_{\ell}
\nonumber\\
&&\hspace{-0.65cm}
=\min\Big\{P_{\max},\eta^*(0)\frac{N_0}{\alpha_{\ell}|h_{\ell}|^2}\Big\}T_{\text{OFDM}}QG - \min\Big\{P_{\max},
\nonumber\\
&&\hspace{-0.65cm}
\eta^*\!\left(\widetilde{h}_{\ell},a_{\ell}\right)\frac{N_0}{\alpha_{\ell}|h_{\ell}|^2}\Big\}T_{\text{OFDM}}QG
\!-\! (P_r\!\!-\!\!P_s)T_{\text{OFDM}}Q(G\!-\!1), \nonumber
\end{eqnarray}
where $\eta^*(0)$ denotes the minimum UL SNR required for traditional forward channel codes to achieve the target PER.

We denote by $\eta_{\ell}^{\text{max}}\triangleq \frac{\alpha_{\ell}|h_{\ell}|^2}{N_0}P_{\max}$ the maximum UL received SNR at the AP when the IoT device transmits with the maximum power $P_{\max}$, and $\eta^{\text{max}}_{\ell,\text{dB}}$ the decibel form of $\eta_{\ell}^{\text{max}}$.
Multiplying the both sides of \eqref{eq:APPC2} by $\frac{\alpha_{\ell}|h_{\ell}|^2}{N_0T_{\text{OFDM}}QG}$, we have
\begin{eqnarray}\label{eq:APPC3}
&&\hspace{-0.65cm}
H\triangleq\frac{\alpha_{\ell}|h_{\ell}|^2}{N_0T_{\text{OFDM}}QG}\left[\text{E}^\prime_{\ell}(a_{\ell})-\text{E}^\prime_{\ell}(0)\right]
\\
&&\hspace{-0.65cm}
=\min\Big\{\eta_{\ell}^{\text{max}},\eta_{\ell}^*(0)\Big\} - \min\Big\{\eta_{\ell}^{\text{max}},\eta^*\left(\widetilde{h}_{\ell},a_{\ell}\right)\Big\}
\nonumber\\
&&\hspace{-0.3cm}
- \frac{G-1}{G}(P_r-P_s)\frac{\alpha_{\ell}|h_{\ell}|^2}{N_0}. \nonumber
\end{eqnarray}

To further simplify \eqref{eq:APPC3}, we have to compare $\eta_{\ell}^{\text{max}}$, $\eta_{\ell}^*(0)$, and $\eta^*\left(\widetilde{h}_{\ell},a_{\ell}\right)$. 

Case I: $\eta_{\ell}^*(0)\geq \eta_{\ell}^{\text{max}}$ and $\eta^*\left(\widetilde{h}_{\ell},a_{\ell}\right)\geq  \eta_{\ell}^{\text{max}}$. In this case,
\begin{equation*}
H= \eta_{\ell}^{\text{max}} - \eta_{\ell}^{\text{max}} - \frac{G-1}{G}(P_r-P_s)\frac{\alpha_{\ell}|h_{\ell}|^2}{N_0} < 0.
\end{equation*}
Therefore, $\text{E}^\prime_{\ell}(a_{\ell})<\text{E}^\prime_{\ell}(0)$ and $\mu(\bm{s}_{\ell},a_{\ell})<0$.

Case II: $\eta_{\ell}^*(0)< \eta_{\ell}^{\text{max}}$ and $\eta^*\left(\widetilde{h}_{\ell},a_{\ell}\right)\geq  \eta_{\ell}^{\text{max}}$. In this case,
\begin{equation*}
H= \eta_{\ell}^*(0) - \eta_{\ell}^{\text{max}} - \frac{G-1}{G}(P_r-P_s)\frac{\alpha_{\ell}|h_{\ell}|^2}{N_0} < 0.
\end{equation*}
Therefore, we also have $\text{E}^\prime_{\ell}(a_{\ell})<\text{E}^\prime_{\ell}(0)$ and $\mu(\bm{s}_{\ell},a_{\ell})<0$.

The common thread in the first two scenarios is that $\eta^*\left(\widetilde{h}_{\ell},a_{\ell}\right)\geq \eta_{\ell}^{\text{max}}$, indicating that the UL received SNR required by DEEP-IoT, $\eta^*\left(\widetilde{h}_{\ell},a_{\ell}\right)$, exceeds what the UL link can possibly provide, even at the maximum transmission power. Consequently, it is advantageous to opt for traditional forward error correction coding rather than feedback coding under these conditions. This rationale underpins the result that $\mu(\bm{s}_{\ell},a_{\ell})<0$.

Case III: $\eta_{\ell}^*(0)\geq \eta_{\ell}^{\text{max}}$ and $\eta^*\left(\widetilde{h}_{\ell},a_{\ell}\right)<  \eta_{\ell}^{\text{max}}$. In this case,
\begin{equation*}
H= \eta_{\ell}^{\text{max}} - \eta^*\left(\widetilde{h}_{\ell},a_{\ell}\right) - \frac{G-1}{G}(P_r-P_s)\frac{\alpha_{\ell}|h_{\ell}|^2}{N_0}.
\end{equation*}
Thus, $H\leq 0$ if $\eta^*\left(\widetilde{h}_{\ell},a_{\ell}\right)\geq\eta_{\ell}^{\text{max}}  -\frac{G-1}{G}(P_r\!-\! P_s)\allowbreak\frac{\alpha_{\ell}|h_{\ell}|^2}{N_0}$.

% However, from the conditions, we know that $\eta^*\left(\widetilde{h}_{\ell},a_{\ell}\right)<\eta^{\text{max}}_{\text{dB}}\leq\eta^*(0)$. This suggests that the target PER can be achieved by only feedback coding, but not the traditional forward coding. Thus, although the sign of $\mu(\bm{s}_{\ell},a_{\ell})<0$ is undetermined, we should choose $a_{\ell}$ in case III.

Case IV: $\eta_{\ell}^*(0)< \eta_{\ell}^{\text{max}}$ and $\eta^*\left(\widetilde{h}_{\ell},a_{\ell}\right)<  \eta_{\ell}^{\text{max}}$. In this case,
\begin{equation*}
H= \eta_{\ell}^*(0) - \eta^*\left(\widetilde{h}_{\ell},a_{\ell}\right) - \frac{G-1}{G}(P_r-P_s)\frac{\alpha_{\ell}|h_{\ell}|^2}{N_0},
\end{equation*}
Thus, $H\leq 0$ if $\eta^*\left(\widetilde{h}_{\ell},a_{\ell}\right)\geq\eta_{\ell}^*(0)  -\frac{G-1}{G}(P_r\!-\! P_s)\allowbreak\frac{\alpha_{\ell}|h_{\ell}|^2}{N_0}$.

Combing the four cases, $H\leq 0$ and $\mu(\bm{s}_{\ell},a_{\ell})\leq 0$ if and only if
\begin{equation*} \eta^*\left(\widetilde{h}_{\ell},a_{\ell}\right)\geq\min\left\{\eta_{\ell}^*(0),\eta_{\ell}^{\text{max}} \right\} -\frac{G-1}{G}(P_r\!-\! P_s)\allowbreak\frac{\alpha_{\ell}|h_{\ell}|^2}{N_0}.
\end{equation*}
The second part of Theorem \ref{thm:1} is proved.
\end{NewProof}

\begin{rem}\label{rem:2}
The monotonicity of the indexes $\mu(\bm{s}_{\ell},a_{\ell})$, $\forall a_{\ell}> 0$, reveals a preference within the single-device MDP framework for the AP to either allocate all $M$ subcarriers to the device or limit the allocation to just one subcarrier. However, it is crucial to highlight that in scenarios where the DL channel SNR $\widetilde{\eta}_{\text{dB}}< -\frac{u_1}{u_2}$, the DL channel SNR is so low that the performance of feedback codes is often inferior to that of forward error correction codes, such as Polar or Turbo codes. Under these conditions, we typically have $\mu(\bm{s}_{\ell},a_{\ell})<0$, making it more advantageous not to allocate any subcarriers to the IoT device. 
As such, the scenario where $\widetilde{\eta}_{\text{dB}}< -\frac{u_1}{u_2}$ becomes irrelevant for DEEP-IoT, focusing our attention on conditions where $\widetilde{\eta}_{\text{dB}}\geq -\frac{u_1}{u_2}$. This delineation ensures that the indexes, $\mu(\bm{s}_{\ell},a_{\ell})$, would maintain a monotonically non-decreasing pattern if allocation decisions are made within the viable SNR region.  
\end{rem}

With the implementation of the action index, we have established a metric to quantify the AP's readiness to allocate resources to each IoT device. Leveraging this concept, we proceed to introduce our index table-based policy gradient (ITPG) algorithm.

At the start of cycle $t$, the IoT cell finds itself in state $\bm{s}^{(t)} = \big\{\bm{s}^{(t)}_{\ell}: \forall \ell  \big\}$ and our goal is to determine $\bm{a}^{(t)}= \big\{a^{(t)}_{\ell}: \forall \ell  \big\}\in\mathcal{A}$. 
We first parameterize the index function $\mu$ by a DNN $\Omega_{\omega}$ with a vector of weights $\omega$. When given the state of a device $\bm{s}^{(t)}_{\ell}$ as the input, $\Omega_{\omega}$ generates the indexes for all possible actions $a^{(t)}_{\ell}$:
\begin{equation}
   \mu\Big(\bm{s}^{(t)}_{\ell},a^{(t)}_{\ell}\Big) = \Omega_{\omega}\Big(\bm{s}^{(t)}_{\ell}\Big),~\forall a^{(t)}_{\ell}.
\end{equation}

With $\Omega_{\omega}$, we compute indexes $\mu(\bm{s}^{(t)}_{\ell},a^{(t)}_{\ell})$ for each user $\ell=1,2,...,L$ and their possible actions $a^{(t)}_{\ell}=0,1,2,...,M$.
Given a total budget of $M$ subcarriers, we can construct an index table $\mu_*$ for $\bm{a}^{(t)}= \big\{a^{(t)}_{\ell}: \forall \ell  \big\}\in\mathcal{A}$, and select the action with the largest index. Specifically, the index of $\bm{a}^{(t)}$ measures the total benefits that can be gained by executing $\bm{a}^{(t)}$. Thus, it is a sum of the indexes of each IoT device when assigning $\bm{a}^{(t)}_{\ell}$ subcarriers to them, giving
    \begin{equation}
\mu_*(\bm{s}^{(t)},\bm{a}^{(t)})\triangleq\sum_{\ell=1}^{L} \mu(\bm{s}^{(t)}_{\ell},a^{(t)}_{\ell}).
    \end{equation}
where $\bm{s}^{(t)}_{\ell}\subset \bm{s}^{(t)}$ and $a^{(t)}_{\ell}\subset \bm{a}^{(t)}$.
The action $\big(\bm{a}^{(t)}\big)^*$ that has the maximum index is denoted by
\begin{equation}
\big(\bm{a}^{(t)}\big)^*=\arg\max_{\bm{a}^{(t)}} \mu_*(\bm{s}^{(t)},\bm{a}^{(t)}).
\end{equation}

The number of entries in the index table $\mu_*$ is the cardinality of the action space $\mathcal{A}$. However, according to Theorem \ref{thm:1}, some entries can be excluded from comparison.

\begin{cor}\label{thm:cor1}
When the IoT cell is in state $\bm{s}^{(t)} = \big\{\bm{s}^{(t)}_{\ell}: \forall \ell  \big\}$, a subcarrier allocation $\widehat{\bm{a}}^{(t)}= \big\{\widehat{a}^{(t)}_{\ell}: \forall \ell  \big\}$ does not need to be considered when constructing the index table $\mu_*$, if a device $\ell$ satisfies either of the following three conditions:
\begin{enumerate}
\item $\eta^*\!\left(\widetilde{h}_{\ell},M\right)\geq  \eta_{\ell}^{\text{max}}$ and $\widehat{a}^{(t)}_{\ell}\geq 1$.
\item $\eta_{\ell}^*(0)\leq \eta_{\ell}^{\text{max}}$ and $1\leq\widehat{a}^{(t)}_{\ell} \leq \lfloor a_{\text{th}} \rfloor$, where
\begin{equation*}
\hspace{-0.6cm}a_{\text{th}}=\frac{1}{u_1\!+\!u_2\widetilde{\eta}_{\ell,\text{dB}}}
\ln\left(\frac{1}{\eta'_{\ell,\text{dB}}-u_5}-u_4\right)-
\frac{u_3\!+\!u_0\widetilde{\eta}_{\ell,\text{dB}}}{u_1\!+\!u_2\widetilde{\eta}_{\ell,\text{dB}}},
\end{equation*}
and $\eta'_{\ell,\text{dB}}=10\lg \left(\eta^{*}_{\ell}(0)-\frac{G-1}{G}(P_r\!-\! P_s)\allowbreak\frac{\alpha_{\ell}|h_{\ell}|^2}{N_0} \right)$.
\item $\eta_{\ell}^*(0)> \eta_{\ell}^{\text{max}}$ and $1\leq\widehat{a}^{(t)}_{\ell} \leq \lfloor a^\prime_{\text{th}} \rfloor$, where
\begin{equation*}
\hspace{-0.6cm}a^\prime_{\text{th}}=\frac{1}{u_1\!+\!u_2\widetilde{\eta}_{\ell,\text{dB}}}
\ln\left(\frac{1}{\eta^{\text{max}}_{\ell,\text{dB}}-u_5}-u_4\right)-
\frac{u_3\!+\!u_0\widetilde{\eta}_{\ell,\text{dB}}}{u_1\!+\!u_2\widetilde{\eta}_{\ell,\text{dB}}}.
\end{equation*}
\end{enumerate}
\end{cor}

\begin{NewProof}
(sketch)
We consider the four cases in the proof of Theorem \ref{thm:1}. In the first two cases, we have 
$\eta^*\!\left(\widetilde{h}_{\ell},a_{\ell}\right)\geq  \eta_{\ell}^{\text{max}}$.
From the monotonicity of the index and comment \ref{rem:2}, it is known that the index is monotonically non-decreasing. Therefore, if $\eta^*\!\left(\widetilde{h}_{\ell},M\right)\geq  \eta_{\ell}^{\text{max}}$, then no action of the IoT device can achieve a positive index. This suggests that no subcarriers should be allocated to the IoT device, which constitutes the first condition in the corollary.

Then, for case IV, where $\eta_{\ell}^*(0)< \eta_{\ell}^{\text{max}}$ and $\eta^*\left(\widetilde{h}_{\ell},a_{\ell}\right)<  \eta_{\ell}^{\text{max}}$, both traditional forward channel coding and DEEP-IoT can meet the receiver's PER requirement. In this case, we have to find the critical action $a_{\text{th}}$ from which the index $\mu$ changes from negative to positive. This can be obtained by setting $H=0$. The final critical value is given in the third condition of the corollary.

Finally, for case III, where $\eta_{\ell}^*(0)\geq \eta_{\ell}^{\text{max}}$ and $\eta^*\left(\widetilde{h}_{\ell},a_{\ell}\right)<  \eta_{\ell}^{\text{max}}$, traditional forward channel coding can no longer meet the receiver's PER requirement, and we must use DEEP-IoT for transmission. Let $\eta_{\ell}^{\text{max}}=\eta^*\left(\widetilde{h}_{\ell},a_{\ell}\right)$, we can obtain the critical action $a^\prime_{\text{th}}$, as indicated by the second condition in the corollary.
\end{NewProof}

\begin{rem}
The general principle of Corollary \ref{thm:cor1} is to exclude actions that yield negative indexes for any IoT device, because allocating feedback resources to them would be less beneficial than not allocating at all.
That said, not all actions with indexes $\mu(\bm{s}^{(t)}_{\ell},a^{(t)}_{\ell})\leq 0$, $\forall a^{(t)}_{\ell}$ should be excluded from consideration. This is because, under certain conditions, traditional forward channel coding may fail to achieve the target PER at the AP. In such scenarios, the only viable option is DEEP-IoT. Essentially, the conditions given in Corollary \ref{thm:cor1} serve as a preliminary filtering process. They preemptively eliminate devices that do not qualify for subcarrier allocation, thereby simplifying the complexity involved in the decision-making process of the index table.
\end{rem} 

% Overall, the index acts as a measurement of how much the AP is willing to pay to allocate $\bm{a}^{(t)}$ subcarriers to the IoT devices. In the beginning of a cycle, we only need to compute and compare the allocating cost across all devices, the optimal subcarrier allocation solution can be found.
% This method not only alleviates the computational burden but also streamlines the policy optimization process, making it a pragmatic solution for managing large-scale IoT environments with enhanced efficiency.

Overall, the ITPG algorithm is summarized in Algorithm \ref{algo:1}.

\begin{algorithm}[t]
\caption{Index table-based policy gradient (ITPG).}
\label{algo:1}
\begin{algorithmic}[1]
\State{Continuously sample trajectories and the update the index DNN $\Omega_{\omega}$.}
\For{each trajectory $\tau$}
\For{$t=1:T(\tau)$}
    \State{Compute the indexes for each device:
\begin{equation*}
   \mu\Big(\bm{s}^{(t)}_{\ell},a^{(t)}_{\ell}\Big) = \Omega_{\omega}\Big(\bm{s}^{(t)}_{\ell}\Big),~\forall \ell.
\end{equation*}}
    \State{Construct an index table $\mu_*$:}
    \If{a pair of $\bm{s}^{(t)}$ and $a^{(t)}$ satisfies the three conditions in Corollary \ref{thm:cor1}}
    \State{$\mu_*\Big(\bm{s}^{(t)}_{\ell},a^{(t)}_{\ell}\Big)=0$.}
    \Else
    \State{
$\mu_*(\bm{s}^{(t)},\bm{a}^{(t)})\triangleq\sum_{\ell=1}^{L} \mu(\bm{s}^{(t)}_{\ell},a^{(t)}_{\ell})$.}
    \EndIf
    \State{$\pi_{\omega}\Big( \bm{a}^{(t)}(\tau)|\bm{s}^{(t)}(\tau) \Big)=\frac{\exp\{\mu_*(\bm{s}^{(t)}(\tau),\bm{a}^{(t)}(\tau))\}}{\sum_{\bm{s}^{(t)},\bm{a}^{(t)}}\exp\{\mu_*(\bm{s}^{(t)},\bm{a}^{(t)})\}}$.}
\EndFor
\State{At the end of $\tau$,
\begin{equation*}
\omega \leftarrow \omega  + \epsilon\nabla_{\omega}\mathbb{E}_{\tau\sim\pi_{\omega }}\big[T(\tau)\big].
\end{equation*}}
\EndFor
\end{algorithmic}
\end{algorithm}

\subsection{Performance}
This section evaluates the efficacy of the ITPG algorithm within the DEEP-IoT MAC layer through simulation. Our simulation environment consists of an IoT cell with $L$ devices, and the number of feedback subcarriers is matched to the number of devices, i.e., $M=L$. The DEEP-IoT physical layer configuration adheres to the parameter settings outlined in Table~\ref{tab:params}.

To address the feedback channel allocation challenge in \eqref{eq:P1}, we employ both the DP with function approximation method and the ITPG approach. These methods are utilized to derive channel allocation policies for DEEP-IoT, with the aim of optimizing the average lifespan of the IoT cell, measured in the maximum number of cycles the IoT cell can operate before one device depletes its battery.

In implementing the ITPG algorithm, we introduce two enhancements to the optimization of the index network $\Omega_{\omega}$ i.e., the last row of Algorithm \ref{algo:1}:
\begin{itemize}
\item We adopt a state-dependent instantaneous reward function $r^{(t)}=\frac{T}{\rho^T-1}(\rho^{t}-\rho^{t-1})$, where $\rho=1.07$ is a predefined constant, in place of the standard state-independent reward $r^{(t)}=1$.
\item We enhance the optimization process by substituting the gradient ascent method with the Proximal Policy Optimization (PPO) algorithm.
\end{itemize}
These advancements facilitate more stable and efficient policy updates, thereby improving the overall performance of the channel allocation process.

\begin{figure}[t]
  \centering
  \includegraphics[width=0.8\columnwidth]{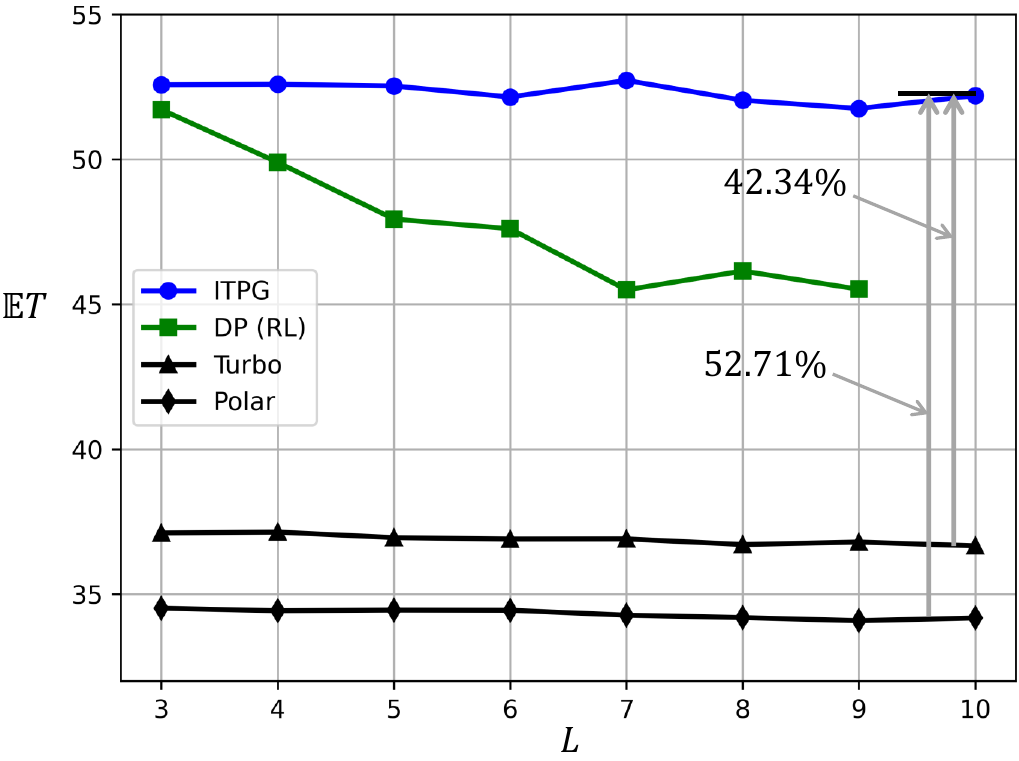}\\
  \caption{The average lifespan $T$ (\# of cycles) of an IoT cell with $L$ devices when operated with DEEP-IoT, benchmarked against traditional forward communication with Turbo and Polar codes.}
\label{fig:cell}
\end{figure}

The simulation results, depicted in Figure \ref{fig:cell}, benchmark the performance of DEEP-IoT (with DP and ITPG) against traditional forward communication methods employing Turbo and Polar codes. 

\begin{figure}[t]
  \centering
  \includegraphics[width=1\columnwidth]{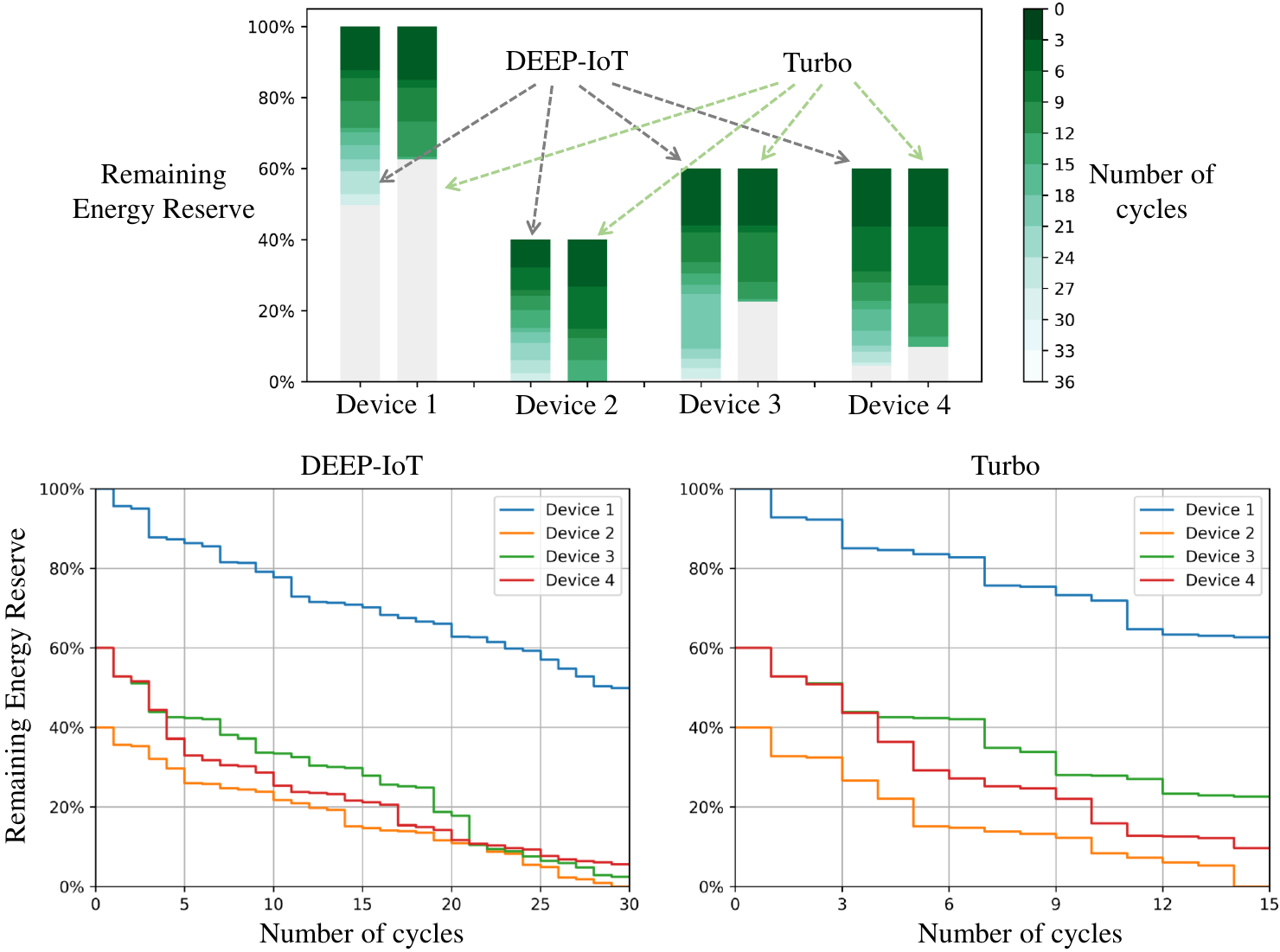}\\
  \caption{The energy consumption of each device over the lifespan of the IoT cell with operated with DEEP-IoT and Turbo codes, where $L=4$.}
\label{fig:cell2}
\end{figure}

First, we juxtapose the efficacy of DP with function approximation against the ITPG algorithm. The results illustrate that for a smaller number of IoT devices $L$, both ITPG and DP with function approximation deliver comparable average lifespans for the IoT cell. However, as $L$ increases, the performance of DP with function approximation deteriorates markedly due to the rising complexity associated with identifying the optimal decision -- a complexity that escalates exponentially with $L$. Conversely, ITPG's performance remains stable, unaffected by the increase in $L$, a testament to the efficiency of the index table-based approach in reducing the dimensionality of the MDP. Particularly, at $L=10$, where the action space cardinality surges to $1.85\times 10^5$, DP becomes impractical, whereas ITPG continues to excel.

When focusing on traditional forward communication utilizing Turbo and Polar codes, the average lifespan of the IoT cell diminishes as $L$ grows. This is because, as the number of devices increases, the probability that at least one device experience deep fade -- and thus deplete its battery prematurely -- increases, adversely affecting the average lifespan of the IoT cell. Despite this, as depicted in Fig. \ref{fig:cell}, the observed performance reduction is relatively minor. Yet, the notable performance disparity between the ITPG method and traditional forward communication techniques underscores the considerable efficiency gains achievable with DEEP-IoT. Specifically, when employing ITPG, DEEP-IoT significantly enhances the IoT cell's performance, achieving a $42.34\%$ improvement over traditional operations using Turbo codes, and a $52.71\%$ enhancement over those with Polar codes.

Considering a scenario with $L=4$ devices, Fig.~\ref{fig:cell2} displays the energy consumption of each device over the lifespan of the IoT cell. The simulation results clearly demonstrates how DEEP-IoT, through strategic allocation of feedback channel resources to devices with lower energy reserves, effectively balances the energy consumption across different devices. Compared to a traditional IoT cell utilizing Turbo codes, the DEEP-IoT system effectively doubles the lifespan of the cell.

\section{Conclusion}\label{sec:Conclusion}
In this work, we have unveiled DEEP-IoT, a groundbreaking shift in IoT communication paradigms that veers away from the conventional transmitter-centric approaches. By pivoting to a receiver-focused model and advocating for a ``listen more, transmit less" philosophy, DEEP-IoT disrupts traditional practices, establishing a novel standard for energy efficiency and device longevity within the IoT domain.

% This initiative is born from the recognition that IoT communications, historically aligned with the unidirectional models of cellular communications, require a fresh perspective that acknowledges their unique operational needs:
% \begin{itemize}
%     \item The insensitivity of IoT communications to latency, pivotal in applications where data transmission is infrequent yet critical for device longevity.
%     \item The typically concise nature of IoT data packets, which presents an opportune landscape for deep learning to derive efficient feedback coding structures.
% \end{itemize}

The ramifications of DEEP-IoT are profound. The exploration in this paper is merely the commencement of a broader inquiry. DEEP-IoT represents a pivotal step forward in our quest for a more sustainable and efficient IoT paradigm. By embracing the principles of device longevity, DEEP-IoT not only offers a viable solution to the pressing challenges facing current IoT systems but also paves the way for the next wave of innovations in IoT communications, from satellite communications to smart healthcare, from infrastructure monitoring to intelligent agriculture and beyond. As we continue to explore and refine this promising paradigm, DEEP-IoT stands to play a crucial role in shaping the future of the Internet of Things.

\appendices

\section{DNN Architectures and Training Methods}\label{sec:AppD}
\subsection{DNN architectures}
\begin{figure}[t]
  \centering
  \includegraphics[width=1\columnwidth]{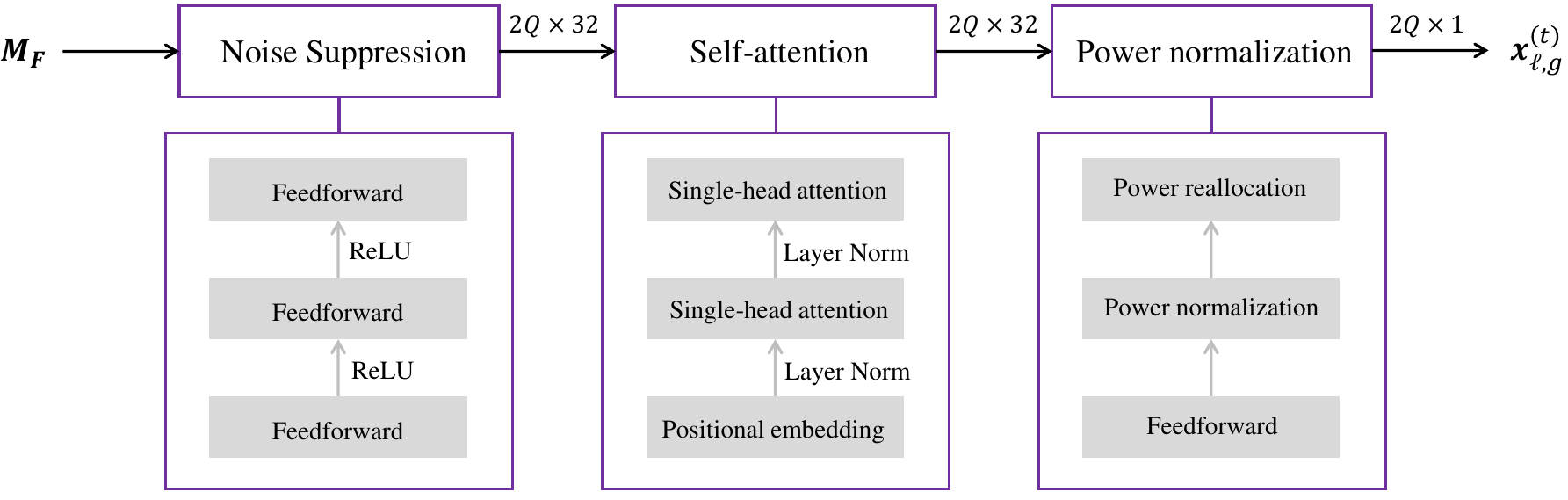}\\
  \caption{The DNN architecture of the uplink feedback encoder at the IoT devices.}
\label{fig:DNNenc}
\end{figure}

The UL feedback encoder $f_F$ comprises three key components: noise suppression, self-attention, and power normalization, as shown in Fig. \ref{fig:DNNenc}.
\begin{itemize}[leftmargin=0.35cm]
    \item The noise suppression module is constructed using three fully connected layers coupled with two ReLU activation functions. Its primary function is to clamp the received signals, effectively reducing excessively large absolute values, which are introduced by noise \cite{GBAFC}. This process ensures that the integrity of the signal is maintained by mitigating noise.
    \item The self-attention mechanism, a key component of feedback encoding, utilizes stacked encoders from the transformer architecture \cite{transformer}. This module excels in selectively concentrating on different parts of the feature matrix, facilitating a dynamic and context-sensitive analysis of feedback. Essentially, it intelligently assesses the AP's current decoding status using the feedback information and adaptively refines its encoding strategy to rectify any discrepancies at the AP's end.
    \item The output of the self-attention mechanism then proceeds to the power normalization module \cite{DeepCode}. This module is responsible for normalizing and redistributing power among the coded channel symbols, ensuring they adhere to the average power constraints.
\end{itemize}

The central element of the feedback encoder is the single-head attention layer, which correlate the encoding of $2Q$ input groups. To provide an analogy from coding theory, these $2Q$ input groups function as variable nodes, while the generated coded symbols in $\bm{x}^{(t)}_{\ell,g}$ act as check nodes. During each encoding cycle, the encoder prioritizes variable nodes that have not been successfully decoded, generating new variable nodes based on a learned degree distribution. This allows the decoder to correct any previously unsuccessful decoding attempts.

The DL feedback encoder $f_B$ and the decoder $f_D$ are architecturally similar to $f_F$, as described in Section \ref{sec:IVB}.

\subsection{Training methods}
Throughout the training process, we maintain a learning rate of 0.001, utilize a batch size of 4096, and implement a weight decay of 0.01. To effectively manage the challenges posed by extensive and very noisy feedback, we adopt a curriculum learning approach, structured into three distinct phases totaling 120,000 training steps.
\begin{itemize}
    \item \textbf{Phase 1:} The initial phase of training starts with a forward SNR of 3dB and a feedback SNR treated as infinite (noiseless feedback). This phase spans 40,000 steps and is designed to stabilize the initial learning by providing noiseless feedback.
    \item \textbf{Phase 2:} During the second phase, lasting 50,000 steps, we introduce data augmentation to tackle the diminishing learning progress often observed as the model’s loss approaches zero. Traditional methods of adjusting loss through hyperparameters would contradict our objective of maintaining specific SNR levels. Therefore, we apply random perturbations to both forward and feedback SNR levels, modeled as a normal distribution with a mean of 0 and a standard deviation of 1.5. This method introduces a controlled variability that increases the incidence of extreme noise instances, enhancing the model's robustness without deviating from the average SNR.
    \item \textbf{Phase 3:} The final phase focuses on training under the specific SNR conditions that the model will encounter. This phase comprises the last 30,000 steps and is crucial for fine-tuning the model’s performance under the designated noisy feedback and feedforward conditions.
\end{itemize}

In addition to the curriculum learning framework, we use additional training techniques:
\begin{itemize}
\item \textbf{Dropout:} We implement dropout at the final layer of the decoder. Contrary to previous studies which suggested negative impacts of dropout on the training of deep learning-based feedback codes, we observe beneficial effects. Specifically, a dropout rate of 0.03 is applied, simulating the training of multiple network configurations and thus enhancing robustness in the face of extensive and noisy feedback.
\item \textbf{Transfer Learning:} Recognizing that some models exhibit superior performance, we employ transfer learning to capitalize on these successful models. By using them as foundational models, we accelerate the development of new models tailored to specific channel conditions. This approach not only shortens the training duration but also enhances performance across various SNR settings, leveraging the pre-trained models’ learned coding structures.
\end{itemize}

Collectively, the above training strategies enhance DEEP IoT's ability to function effectively in extensive feedback and very noisy feedback scenarios.

\section{Proof of Lemma \ref{thm:lem1}}\label{sec:AppA}
For the state $\bm{s}_{\ell}$, we have 
\begin{eqnarray}\label{eq:APPA1}
v_*(\bm{s}_{\ell})
\hspace{-0.2cm}&=&\hspace{-0.2cm}
\mathbb{E}_{\tau\sim\pi_*}\left[\sum_{t=1}^{T(\tau)}r_{\ell}^{(t)}\Bigg|\bm{s}_{\ell}^{(1)}=\bm{s}_{\ell}\right] \nonumber\\
\hspace{-0.2cm}&=&\hspace{-0.2cm}
\int_{\big\{h^{(t)}_{\ell}\!(\tau),\widetilde{h}^{(t)}_{\ell}\!(\tau),a^{(t)}_{\ell}\!(\tau)\big\}_{t=1}^{T(\tau)}} T(\tau)\Pr(\tau)d\tau,
\end{eqnarray}
where 
$$\Pr(\tau)\!=\!\sum_{t=1}^{T(\tau)}\pi_{*}\Big( a_{\ell}^{(t)}(\tau)|s_{\ell}^{(t)}(\tau) \Big)\Pr\Big(h^{(t+1)}_{\ell}(\tau),\widetilde{h}^{(t+1)}_{\ell}(\tau)\Big).$$
To simply notation, we define $e_{\tau}(1,T(\tau))\triangleq \big\{h^{(t)}_{\ell}(\tau),\allowbreak\widetilde{h}^{(t)}_{\ell}\allowbreak(\tau),\allowbreak a^{(t)}_{\ell}(\tau)\big\}_{t=1}^{T(\tau)}$.

For the state $\bm{s}^{\prime}_{\ell}$, we construct a new policy $\pi^\prime$: 
$$a^{(t)}_{\ell}\!(\tau)=\pi^\prime\Big(h^{(t)}_{\ell}\!(\tau),\widetilde{h}^{(t)}_{\ell}\!(\tau)\Big),$$ 
which produces actions based only on the UL and DL channel realizations. In particular, the behavior of $\pi^\prime$ matches exactly with the trajectory $\tau$ in \eqref{eq:APPA1}, and
\begin{equation}\label{eq:APPA0}
    v_{\pi^\prime}(\bm{s}^{\prime}_{\ell}) \leq v_*(\bm{s}^{\prime}_{\ell}).
\end{equation}

Following \eqref{eq:APPA1}, we have
\begin{eqnarray}\label{eq:APPA2}
&&\hspace{-0.5cm}
v_{*}(\bm{s}_{\ell}) - v_{\pi^\prime}(\bm{s}^{\prime}_{\ell})
 \\
&&\hspace{-0.5cm}=
\int_{e_{\tau}(1,T(\tau))} T(\tau)\Pr(\tau)d\tau - \int_{e_{\tau^\prime}(1,T(\tau^\prime))} T(\tau^\prime)\Pr(\tau^\prime)d\tau^\prime. \nonumber
\end{eqnarray}
Notice that the UL and DL channels are independent of the actions, and the policy $\pi^\prime$ is designed such that $a^{(t)}_{\ell}\!(\tau)=a^{(t)}_{\ell}\!(\tau^\prime)$ under the same channel conditions. 
There must exist a critical cycle $T_c$, for which 1) $e_{\tau}(1,T_c)=e_{\tau^\prime}(1,T_c)$, and 2) either $\text{E}_{\ell}^{(T_c+1)}(\tau)<0$ or $\text{E}_{\ell}^{(T_c+1)}(\tau^\prime)<0$.

It can be verified from \eqref{eq:P1b} that the energy expenditures of $\tau$ and $\tau^\prime$ in the first $T_c$ cycles are exactly the same. Therefore,
\begin{equation*}
    \text{E}_{\ell}^{(T_c+1)}(\tau^\prime)-\text{E}_{\ell}^{(T_c+1)}(\tau)=\text{E}_{\ell}^{(1)}(\tau^\prime)-\text{E}_{\ell}^{(1)}(\tau)=\text{E}^\prime_{\ell}-\text{E}_{\ell}\geq 0.
\end{equation*}
This suggests that $\tau$ enters the terminal state after $T_c$ cycles. We can then partition the trajectory $\tau^\prime$ into two phases: $\tau^\prime=[\tau,\tau_c]$, where the number of cycles in $\tau$ is $T_c$.

The second term on the right hand side (RHS) of \eqref{eq:APPA2} can be refined as
\begin{eqnarray}\label{eq:APPA3}
&&\hspace{-0.8cm}
\int_{e_{\tau^\prime}(1,T(\tau^\prime))} T(\tau^\prime)\Pr(\tau^\prime)d\tau^\prime
 \\
&&\hspace{-0.8cm} = \!\int_{e_{\tau}(1,T(\tau))} \!\int_{e_{\tau^\prime}(T(\tau)+1,T(\tau^\prime))}\!\!\Big[T(\tau)\!+\!T(\tau_c)\Big]\Pr(\tau)\Pr(\tau_c)d\tau^\prime    \nonumber \\
&&\hspace{-0.8cm} =\int_{e_{\tau}(1,T(\tau))}T(\tau)\Pr(\tau)d\tau  +
\int_{e_{\tau^\prime}(1,T(\tau^\prime))} T(\tau_c) \Pr(\tau^\prime)d\tau^\prime.  \nonumber
\end{eqnarray}

Substituting \eqref{eq:APPA3} into \eqref{eq:APPA2} yields
\begin{equation}\label{eq:APPA4}
v_{*}(\bm{s}_{\ell}) - v_{\pi^\prime}(\bm{s}^{\prime}_{\ell}) = -\int_{e_{\tau^\prime}(1,T(\tau^\prime))} T(\tau_c) \Pr(\tau^\prime)d\tau^\prime \leq 0.
\end{equation}

Combining \eqref{eq:APPA0} and \eqref{eq:APPA4}, we have
\begin{equation}
v_{*}(\bm{s}_{\ell}) \leq v_{\pi^\prime}(\bm{s}^{\prime}_{\ell}) \leq v_*(\bm{s}^{\prime}_{\ell}). 
\end{equation}
Lemma \ref{thm:lem1} is proved.

% \section{Proof of Lemma \ref{thm:lem2}}\label{sec:AppB}
% \input{AppendixB.tex}

% \section{Proof of Theorem \ref{thm:thm1}}\label{sec:AppC}
% \input{AppendixC.tex}

\bibliographystyle{IEEEtran}
\bibliography{References}

\end{document}